\documentclass[twocolumn, twocolappendix, floatfix, a4paper]{aastex631}

\usepackage[caption=false]{subfig}
\usepackage{enumitem}
\usepackage[flushmargin]{footmisc}

\newcommand{\msun}{\, \mathrm{M_{\odot}}}

\shorttitle{Ram pressure stripping in EAGLE}
\shortauthors{Kulier et al.}
\begin{document}

\title{Ram pressure stripping in the EAGLE simulation}

\correspondingauthor{Andrea Kulier}
\email{andrea.kulier@inaf.it}

\author[0000-0002-0431-2445]{Andrea Kulier}
\affiliation{INAF-Osservatorio Astronomico di Padova, Vicolo dell'Osservatorio 5, 35122 Padova, Italy}

\author[0000-0001-8751-8360]{Bianca Poggianti}
\affiliation{INAF-Osservatorio Astronomico di Padova, Vicolo dell'Osservatorio 5, 35122 Padova, Italy}

\author[0000-0002-8710-9206]{Stephanie Tonnesen}
\affiliation{Flatiron Institute, CCA, 162 5th Avenue, New York, NY 10010, USA}

\author[0000-0001-5303-6830]{Rory Smith}
\affiliation{Departamento de F\'isica, Universidad T\'ecnica Federico Santa Mar\'ia, Avenida Vicu\~na Mackenna 3939, San Joaqu\'in, Santiago, Chile}

\author[0000-0003-1581-0092]{Alessandro Ignesti}
\affiliation{INAF-Osservatorio Astronomico di Padova, Vicolo dell'Osservatorio 5, 35122 Padova, Italy}

\author[0000-0001-7011-9291]{Nina Akerman}
\affiliation{INAF-Osservatorio Astronomico di Padova, Vicolo dell'Osservatorio 5, 35122 Padova, Italy}
\affiliation{Dipartimento di Fisica e Astronomia ‘Galileo Galilei’, Universit\'a di Padova, Vicolo dell’Osservatorio 3, 35122 Padova, Italy}

\author[0000-0002-5655-6054]{Antonino Marasco}
\affiliation{INAF-Osservatorio Astronomico di Padova, Vicolo dell'Osservatorio 5, 35122 Padova, Italy}

\author[0000-0003-0980-1499]{Benedetta Vulcani}
\affiliation{INAF-Osservatorio Astronomico di Padova, Vicolo dell'Osservatorio 5, 35122 Padova, Italy}

\author[0000-0002-1688-482X]{Alessia Moretti}
\affiliation{INAF-Osservatorio Astronomico di Padova, Vicolo dell'Osservatorio 5, 35122 Padova, Italy}

\author[0000-0001-5840-9835]{Anna Wolter}
\affiliation{INAF-Osservatorio Astronomico di Brera, via Brera 28, I-20121 Milano, Italy}

%\author{et al.}
%\affiliation{}



\begin{abstract}
Ram pressure stripping of satellite galaxies is thought to be a ubiquitous process 
in galaxy clusters, and a growing number of observations reveal 
satellites at different stages of stripping.
However, in order to determine the fate of any individual galaxy, 
we turn to predictions from either simulations or analytic models. 
It is not well-determined whether simulations and analytic models 
agree in their predictions, nor the causes of disagreement. 
Here we investigate ram pressure stripping in the 
reference EAGLE hydrodynamical cosmological simulation, 
and compare the results to predictions from analytic models.
We track the evolution of galaxies with stellar mass $M_{*} > 10^{9} \msun$
and initial bound gas mass $M_{\mathrm{gas}} > 10^{9} \msun$
that fall into galaxy clusters ($M_{\mathrm{200c}} > 10^{14} \msun$) between
$z = 0.27$ and $z = 0$. We divide each galaxy into its neutral gas disk and hot ionized gas
halo and compare the evolution of the stripped gas fraction
 in the simulation to that predicted by
analytic formulations for the two gas phases, as well as to a toy model
that computes the motions of gas particles under the combined effects of gravity
and a spatially uniform ram pressure. 
We find that the analytic models generally underpredict the stripping rate of neutral gas and overpredict that of ionized gas, with significant scatter between the model and simulation stripping timescales. This is due to opposing physical effects: the enhancement of ram pressure stripping by stellar feedback, and the suppression of stripping by the compaction of galactic gas.

\end{abstract}

%\keywords{}


\section{Introduction} \label{intro}

For decades, it has been known that galaxies in dense regions, such
as galaxy clusters and groups, are more likely 
to be gas-poor and passive than galaxies of the same stellar mass in the field \citep{dressler}.
This difference is attributed to a number of physical processes
prevalent in such regions, including tidal interactions
of satellite galaxies both with other satellites and with the central
cluster potential \citep{byrd, moore},
as well as ram pressure stripping: the removal of gas from satellite galaxies due to 
the pressure created by their motion relative to the
intragroup or intracluster medium \citep{gunngott}. Unlike gravitational interactions, 
ram pressure acts only on the galactic gas, not on the stars. The most
visually spectacular examples of ram pressure stripping are
so-called ``jellyfish'' galaxies, which have undisrupted stellar disks
but long, filamentary gaseous tails \citep{jelly_main}.

An analytic theoretical treatment of the physics of ram pressure stripping
was first provided by \citet{gunngott}. 
Such idealized analytic approximations are often used to estimate
the stripping rate due to ram pressure,
but their assumptions may not be satisfied for real galaxies.
Physical effects that are not captured
by such simplified treatments have been studied using hydrodynamical simulations.
For example, high-resolution simulations of individual galaxies in an oncoming gas flow
have shown that some regions of galactic gas can be shielded
from ram pressure by other regions \citep{roetilt, tilted}. 
Such simulations have also found that ram pressure 
can induce spiral waves 
in the gas that transport angular momentum outward,
leading to the contraction of the inner part of the galaxy and
preventing further stripping
\citep{schulz}. The strength of both of these effects
depends on the orientation of the gas disk relative to the ram pressure direction.

In addition to the \textsc{HI} and molecular gas in the disk, ram pressure
can also remove a galaxy's hot ionized gas halo (i.e. its circumgalactic medium),
depriving the galaxy of its source of accreting gas ---
 a process known as ``strangulation'' \citep{larson}.
Simulations have found that the ionized halo can be stripped by modest
ram pressures, significantly less than what is required to
strip the disk \citep{oldbekki, mccarthy, bekki, steinhauser}.

While high-resolution simulations of individual galaxies allow for 
investigation of physical processes that cannot be captured by lower-resolution 
simulations, they also tend to assume idealized initial conditions for the 
geometry of galaxies and the flow of the intracluster medium (ICM). For example,  
\citet{tonnesenvary} found that a varying ram pressure produces a different stripping rate
than the constant one assumed by many high-resolution simulations.

Cosmological hydrodynamical simulations such as EAGLE \citep{eagleschaye}
and IllustrisTNG \citep{tng} include a wide range of 
galaxy morphologies that fall into dense regions such as groups
and clusters. Due to the cosmological nature of the
simulation, these galaxies should be statistically representative of the true 
variety of galaxies that experience ram pressure.
Galaxies also experience a range of ram pressures 
as they traverse different environments.
However, due to the lower resolution of these simulations, it is possible that
some relevant gas physics is not properly accounted for, 
such as the influence of magnetic fields \citep{ramos}, or the mixing between
the interstellar medium (ISM) and the ICM \citep{mixing, mixing_obs}.

Cosmological hydrodynamical simulations have previously been used to examine
a number of aspects of ram pressure stripping.
In GIMIC \citep{gimic}, a set of simulations of $\sim20 h^{-1}$ Mpc
regions, it was found that stellar feedback enhances the stripping
due to ram pressure \citep{feedback_gimic}, 
while confinement pressure has little effect on it \citep{gimic_confinement}.
\citet{marasco} found that in the EAGLE reference simulation at $z = 0$,
ram pressure stripping was a more common \textsc{HI} gas removal mechanism
acting on galaxies than tidal interactions.
More recently, \citet{pallero} found that in the C-EAGLE simulations \citep{ceagle}, 
a set of resimulations of galaxy clusters using EAGLE physics, the majority
of galaxies that entered the virial radius of a cluster 
became quenched within 1 Gyr, which the authors attribute to ram pressure stripping,
in agreement with a prior cosmological simulation of a single cluster \citep{tonnesen2007}.

In the Illustris TNG100 simulation, \citet{jelly_illustris} examined visually
identified jellyfish galaxies and found that they had higher Mach numbers and were
experiencing higher ram pressures than other galaxies in groups and clusters. They
also found that most jellyfish galaxies were recent infallers, having entered the cluster
virial radius within the previous 3 Gyr. \citet{new_illustris} studied jellyfish 
galaxies in the higher-resolution TNG50 simulation that had 
been visually identified as part of the Galaxy Zoo project \citep{illustris_zoo}.
They similarly found that ram pressure stripping of these galaxies begins within
approximately 1 Gyr of cluster infall and lasts for $\lesssim 2$ Gyr, and that
the majority of stripping occurs between 0.2 and 2 times the cluster virial radius.

Although both high-resolution and cosmological simulations have been used
to theoretically investigate the evolution of ram pressure stripped galaxies, 
simplified analytic models like the one presented in \citet{gunngott} are still
frequently used to estimate the effect of 
ram pressure stripping in the literature --- for example, in semi-analytic
models of galaxy evolution (e.g. \citealt{lgalaxies, threehundred}). 
In this paper, we attempt to assess the differences between the 
fraction of stripped gas predicted by these
models and the results of a full hydrodynamical cosmological simulation, and to analyze
the physical effects missing from the analytic models that give rise to these differences.

Specifically, we use the (100 Mpc)$^3$ reference 
EAGLE cosmological hydrodynamical simulation, in which
we select galaxies with stellar masses
$M_{*} > 10^{9} \msun$ and initial bound gas masses 
of $M_{\mathrm{gas}} > 10^{9} \msun$
that fall into galaxy clusters between the simulation snapshots at 
$z = 0.27$ and $z = 0$, i.e. within the last 3.2 Gyr.
We compare the stripped gas mass in the simulation to that predicted by common analytic prescriptions,
and use the differences between the two to 
determine the additional physical processes that impact
the amount of gas stripped by ram pressure. To assist in doing
the latter, we implement an additional toy model of greater complexity
than the analytic models. This model 
integrates the motion of the gas particles
within the gravitational potential of each subhalo, using the starting 
positions and velocities of the gas particles as initial conditions. 
By comparing this model to both the simulation and simple
analytic models, we are able to separate the different physical effects
that influence the ram pressure stripping rate of simulated galaxies.

In \S\ref{methods}, we provide a brief overview of the EAGLE simulation 
and describe the sample of simulated galaxies used in our analysis.
In \S\ref{models}, we describe the simplified models to which we compare
the results of the simulation. In \S\ref{results}, we present our results
comparing the analytic models for ram pressure stripping, our more complex toy model,
and the EAGLE simulation. In \S\ref{results_toy}, we utilize our toy model
to explain the physical effects that give rise to most of the discrepancies
between the analytic models and the simulation.
Finally, in \S\ref{conclusions}, we summarize our conclusions.

\section{Simulations and Galaxy Sample} \label{methods}

\subsection{EAGLE Simulation Overview} \label{eagle}

EAGLE \citep{eagleschaye, eaglecrain, mcalpine} is a
suite of cosmological hydrodynamical simulations, run using
a modified version \citep{anarchy} of the N-body smooth 
particle hydrodynamics (SPH) code GADGET-3 \citep{gadget}.
These modifications are based on
the conservative pressure-entropy formulation of SPH from \citet{hopkins}, and include
changes to the handling of the viscosity \citep{cullen}, 
the conduction \citep{price}, the smoothing kernel \citep{denhen}, and
the time-stepping \citep{durier}.
EAGLE adopts a Planck cosmology \citep{planck}
with $h = 0.6777$, $\Omega_{\Lambda} = 0.693$, $\Omega_{m} = 0.307$,
and $\Omega_{b} = 0.048$, which we also adopt throughout
this paper.

In this paper we use the reference EAGLE simulation Ref-L0100N1504,
which has a box size of 100 comoving Mpc per side and contains
$1504^{3}$ particles each of dark matter and baryons.
The dark matter particle mass is $9.70\times10^{6} \msun$
and the initial gas (baryon) particle mass is $1.81\times10^{6} \msun$.
The Plummer-equivalent gravitational softening length is 2.66 comoving kpc until $z = 2.8$ and
0.70 proper kpc afterward. Every particle in EAGLE, whether dark matter or baryonic, has a unique
identifier that allows it to be tracked throughout the simulation at different times.

The subgrid physics in EAGLE includes prescriptions for
radiative cooling, photoionization heating,
star formation, stellar mass loss, stellar feedback,
supermassive black hole accretion and mergers, and AGN feedback.
These prescriptions and the effects of varying them are described in \citet{eagleschaye}
and \citet{eaglecrain}.

Radiative cooling and photoionization heating is implemented
using the model of \citet{cooling}. Cooling and heating rates are computed for 11 elements
using CLOUDY \citep{cloudy}, assuming that the gas is optically thin, in ionization equilibrium,
and exposed to the cosmic microwave background and a \citet{hardtmadau} UV and X-ray background.

Star formation is implemented as described in
\citet{sfr}. Gas particles that reach a metallicity-dependent
density threshold \citep{sfr2} become `star-forming' and
are stochastically converted into stars at a rate that
reproduces the Kennicutt-Schmidt law \citep{ks}.
Star particles are modeled as simple
stellar populations with a \citet{chabrier} initial mass function.
Prescriptions for stellar evolution and mass loss from \citet{snwiersma} are
used.

Stellar feedback is modeled using
the stochastic, purely thermal feedback prescription of \citet{supernova}.
When a newly-formed star particle reaches an age of 30 Myr, it injects feedback energy into its surroundings
by heating some number of neighboring gas particles by $10^{7.5}$ K.
The total injected energy 
is calibrated by adjusting the fraction of total stellar feedback energy that heats the nearby gas.
Stellar feedback in the EAGLE reference simulation was calibrated to simultaneously reproduce the local
galaxy stellar mass function and mass-size relation \citep{eaglecrain}.

When a dark matter halo reaches a mass of
$10^{10} \msun/h$, it is seeded with a black hole of subgrid mass
$10^{5} \msun/h$ at its center by converting the most bound gas particle
into a ``black hole'' particle \citep{agn1}. The black hole particles accrete gas
according to the modified Bondi-Hoyle prescription given in \citet{agn2}, and can merge with one another.
AGN feedback is implemented as stochastic and purely thermal,
similar to stellar feedback. The energy injection rate is proportional
to the black hole accretion rate. Gas particles near the 
black hole particle are stochastically heated by $10^{8.5}$ K.
The fraction of lost energy that is assumed to heat the gas does not significantly
affect the masses of galaxies due to self-regulation \citep{booth2010},
and is instead calibrated to match the observed stellar mass-black hole mass relation.

Galaxies are identified in EAGLE through a series of steps.
First, halos are identified in the dark matter particle distribution
using a friends-of-friends (FoF)
algorithm with a linking length of $b = 0.2$ times the mean interparticle separation \citep{davis}.
Other particle types (gas, stars, and black holes) are assigned to
the FoF halo of the nearest dark matter particle.
The \textsc{subfind} \citep{subfind1, subfind2} algorithm is then
run over all the particles of any type within each FoF halo,
in order to identify local overdensities (``subhalos'').
Each subhalo is assigned only the particles
gravitationally bound to it, with no overlap in particles between
subhalos. The subhalo that contains the most bound particle in a FoF halo
is considered to be the ``central'' subhalo, while
any other subhalos are ``satellites''. The gravitationally bound
gas and star particles in each subhalo constitute a ``galaxy''.

As we will describe below, we track
galaxies between $z = 0.27$ and $z = 0$, between which there 
are 28 short-timescale simulation outputs (termed ``snipshots'') 
that have been processed with \textsc{subfind}. These outputs are generally spaced by
$\approx 120$ Myr: in 25 cases they are spaced by between 119 and 128 Myr, 
while two cases are spaced by 59 and 61 Myr, respectively.
The snipshots are useful for studying physical processes that can
be rapid, such as ram pressure stripping.

Galaxy merger trees were created
using the \textsc{d-trees} algorithm \citep{dtrees}
as modified by \citet{qu}.
The merger trees are available at 28 snapshots 
between $z = 20$ and $z = 0$, of which we use four in this paper:
$z = 0.27, 0.18, 0.1$, and 0.

EAGLE has been found to reproduce many observed galaxy properties,
including the $z = 0$ Tully-Fisher relation \citep{eagleschaye}, 
the evolution of the galaxy stellar mass function \citep{furlong2015}, 
the neutral gas masses of galaxies with $10^{10} \msun < M_{*} < 
10^{11} \msun$ \citep{bahehi}, 
the dependence of galaxy \textsc{HI} content on environment \citep{marasco},
the SFR-$M_{*}$ relation \citep{furlong2015}, and the evolution
of the star formation rate function \citep{katsianis2017}.
Galaxy and halo catalogs as well as particle data from EAGLE snapshots have been made
publicly available \citep{mcalpine}.

\begin{deluxetable*}{c| c c c c c}
\tabletypesize{\footnotesize}
\tablecolumns{6}
\tablewidth{0pt}
\tablecaption{Simulated Galaxy Sample Properties \label{table:results1}}
\tablehead{ Infall Redshift & All Gas & Gas Remains & Merges with & Falsely\tablenotemark{a} Merges with & \textbf{Total}\\
& Removed & at $z = 0$ & Cluster Central & Cluster Central & 
}
\startdata
$0 < z < 0.1$ & 2 & 18 & 0 & 0 & 20 \\
$0.1 < z < 0.18$ &10 & 13 & 1 & 1 & 25 \\
$0.18 < z <0.27$ & 18 & 15 & 1 & 1 & 35 \\
\enddata
\tablenotetext{a}{\textsc{Subfind} may sometimes improperly identify two nearby galaxies
as a single object. As a result, the particles of two galaxies in our sample are at one point 
assigned to the central cluster galaxy even though the galaxies reappear as separate objects in
later outputs. Because we cannot track which particles are still bound to a galaxy 
in the outputs in which the particles are improperly assigned to
the central object, we cease tracking the galaxy when this occurs.}
\end{deluxetable*}

\subsection{Simulated Galaxy Sample Selection} \label{galsamp}

We use the EAGLE merger trees to
select a sample of galaxies that fall into clusters between $z = 0.27$ and $z = 0$.
Such galaxies should experience a wide range of ram pressures as they
travel from outside the cluster to the pericenter of their orbits.

In EAGLE there are 7 clusters (FoF groups) with $M_{\mathrm{200c}} > 10^{14} \msun$\footnote{Here 
$M_{\mathrm{200c}}$ is the mass within the virial radius $r_{\mathrm{200c}}$, 
which is the radius within which the mean density
is 200 times the critical density of the Universe, measured centered on the most bound dark matter
particle within each FoF halo.}
at $z = 0$. One of these consists of two
merging clusters (individually more massive than $10^{14} \msun$ at $z = 0.1$) whose central galaxies are still
separated by a distance of approximately $5r_{\mathrm{200c}}$,
and we thus treat the two components as separate clusters.

For each cluster at $z=0$, we identify the FoF group containing the
main progenitor of its central galaxy at each previous output, and
treat this as the same cluster at a previous time. In practice, this is also
the most massive FoF group that merges with others to become the $z = 0$ cluster.

We choose to focus on galaxies that do not 
merge into clusters as satellites of larger structures, in which they may undergo
so-called ``preprocessing'' \citep{preproc}. We defer analysis of a larger sample
of galaxies in a broader range of environments to later work. 
We therefore select our galaxy sample according to the following criteria.
We identify galaxies that at $z = 0.1$, 0.18, or 0.27 were
not part of cluster FoF groups, and were the central galaxies of their own FoF group,
but became members of a cluster FoF group in the
subsequent snapshot. We note that the physical extent of FoF groups is 
typically larger than
$r_{200c}$ and therefore a galaxy may join a cluster
outside of this radius. We also searched for galaxies that were initially
isolated and had merged with the central galaxy of the cluster by the
next snapshot, indicating a rapid descent to the center of the cluster, but found
no such objects.

We enforce cuts on the stellar and gas mass
of our galaxy sample: they must initially have $M_{*} > 10^{9} \msun$
and $M_{\mathrm{gas}} > 10^{9} \msun$. Here
the gas mass is the total mass of gas that
is gravitationally bound to each galaxy subhalo, regardless of whether it is neutral
or ionized; we will discuss the division of gas into different phases
in \S\ref{neutralionized}.

As the purpose of this paper is to investigate ram pressure stripping,
we select those galaxies that lose at least 1/3 of their initial gas mass by
$z = 0$. Because we would like to focus on ram pressure stripping 
and exclude other evolutionary processes that significantly affect galaxies, 
we remove galaxies that the EAGLE merger trees indicate
merge with an object larger than $5\%$ of their stellar or gas mass between
the initial redshift and $z = 0$. Also, we choose only those galaxies whose stellar
mass does not decrease between the two snapshots during which they enter the cluster, 
aside from the mass lost due to the stellar mass loss prescription in EAGLE. This is to avoid
selecting galaxies that are strongly affected by tidal 
interactions, which strip stars as well as gas. However, we do not exclude galaxies that
lose stellar mass later in their evolution through the cluster, or simultaneously
lose stellar mass and gain it via star formation. We analyze the potential
influence of tides on our galaxy sample in Appendix \ref{appendix}, where
we find that tidal interactions (with either other satellite galaxies or the
cluster center) are unlikely to significantly influence
the galaxies in our sample. We conclude that gas stripping
in our galaxy sample is dominated by ram pressure.
 
We identify the selected galaxies in the short-timescale outputs (``snipshots'')
of the EAGLE simulation. We take the descendant of each galaxy
to be the one that contains the majority of its star particles.
For consistency in the sample, we eliminate galaxies
that become satellites of larger groups at any point before cluster infall,
as well as those that increase their baryonic
(stellar plus gas) mass by over $5\%$
at any point after the initial output, which we take to be the
output at which the galaxy has its maximum baryonic mass.

We track these galaxies until one of three conditions is satisfied: the galaxy
loses all of its gas mass, the galaxy merges with another galaxy (in practice,
always the central galaxy of the cluster), or the end of the simulation ($z = 0$) is reached.
For the case of mergers, sometimes \textsc{subfind} falsely ``merges'' two galaxies that
are later identified as separate objects again.
In this analysis, we cease tracking the galaxy
when this occurs, but still retain it as part of our sample as long as its
trajectory is at least 5 simulation outputs
long and it loses at least 33\% of its initial gas mass by that point.
There are two galaxies
in our final sample that falsely ``merge'' with the central galaxy
in this manner and that we cease tracking at that time.

\begin{figure*}
    \centering
    \subfloat{\includegraphics[width=0.48\textwidth]{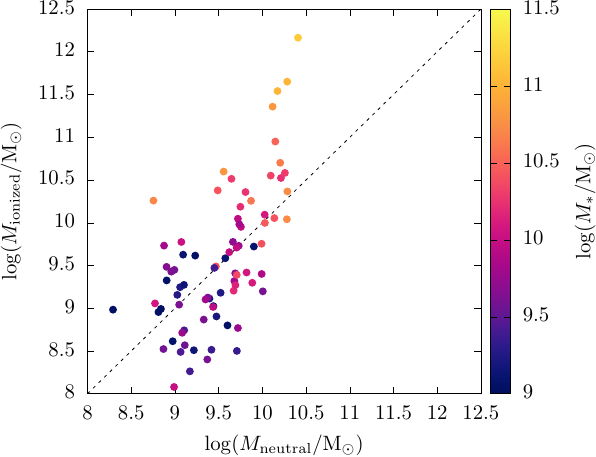}}
    \hfill
    \subfloat{\includegraphics[width=0.48\textwidth]{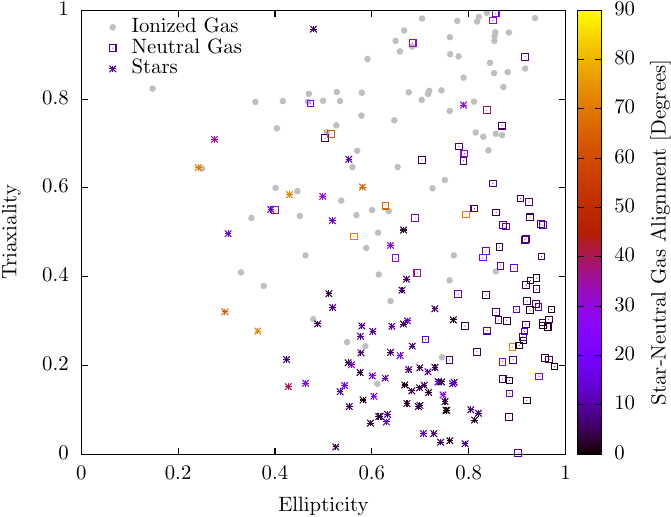}}
    \caption{\label{fig1} Initial properties of our selected ram pressure-affected galaxy sample. \textit{Left:} Neutral gas mass versus ionized gas mass in each galaxy, where the masses are computed as the sum of the masses of the gas particles designated ``neutral'' or ``ionized'' based on their neutral hydrogen fraction (see \S \ref{neutralionized}). The color scale shows the stellar mass of each galaxy. \textit{Right:} The initial morphology of the gas and stellar components of our galaxies, with square points representing the neutral gas, circular points representing the ionized gas, and star-shaped points representing the star particles. The horizontal axis represents the ellipticity $\epsilon = 1 - c/a$, while the vertical axis is the triaxiality $T = (1 - b^{2}/a^{2})/(1 - c^{2}/a^{2})$, where $a \ge b \ge c$ are the axes computed from the reduced moment of inertia tensor of the particle distribution, similar to \citet{thob2019}. Objects to the left side of the plot are spherical whereas objects to the right side are flattened. Objects near the bottom right are oblate disks, whereas those near the top right are prolate. The color scale shows the angular difference between the minor axes of the stellar and neutral gas components of each galaxy, such that 0 degrees indicates alignment.}
\end{figure*}

Selected in this manner, our final sample consists of 80 galaxies.
Their infall redshifts and status at the time we stop tracking
them are summarized in Table \ref{table:results1}.

\subsubsection{Separation of Gas into Neutral and Ionized}
\label{neutralionized}

We separate the gas bound to each galaxy
into a neutral component and a hot ionized component.
One reason to treat these separately is with consideration to observations:
the two phases of gas are not observable in the same bands.
The second reason that is more relevant to the current paper is that
the neutral and ionized gas components have different geometries, with the former
generally located in a flattened, rotating disk and the latter in a spheroidal halo.
The two geometries are treated differently in common analytic
models for ram pressure stripping, which will be described further in \S \ref{analytic}.

EAGLE does not natively track different phases of gas. We instead employ
the prescription used in \citet{bahehi} and \citet{crainhi} to estimate the fraction
of neutral (HI + H$_{2}$) gas versus ionized hydrogen gas
in each gas particle. The prescription is based on 
a fitting formula from \citet{neutral} with
the temperature, pressure, and metallicity of each gas particle as inputs,
and an assumed value of the UV background from \citet{hardtmadau}.
The fitting formula was derived from higher-resolution simulations of galaxies
that implemented radiation transport modeling; details can be found in \citet{neutral}.
When applied to EAGLE galaxies, this prescription results in mean galaxy neutral gas masses
within 0.1 dex of observations for galaxies with $M_{*} > 10^{10} \msun$ at $z = 0$ \citep{bahehi},
although galaxies with lower stellar masses are found to be
HI-deficient compared to observations \citep{crainhi}.

Because we would like to trace the kinematics of individual gas particles, as we will
describe in \S \ref{toymodel}, we do not divide the particles 
into neutral and ionized components, but rather simply label each gas particle as either
``neutral'' or ``ionized'' based on whether its neutral hydrogen fraction
is above or below $50\%$. The distribution
of the neutral hydrogen fraction of gas particles in EAGLE galaxies tends
to be strongly bimodal \citep{manuwal}, and thus the results
should not depend significantly on the choice of $50\%$ as the separation criterion.
Neutral gas particles generally have temperature $T\sim10^{4}$ K, as EAGLE has an imposed temperature
floor due its finite resolution, while ionized gas has $T\sim10^{5}-10^{6}$ K.

\subsection{Galaxy Sample Properties}
\label{samp_properties}

In Figure \ref{fig1}, we present the initial properties of our sample,
separated into ``neutral'' and ``ionized'' gas
as described above. These are the properties of the galaxies in the first simulation output at which
we begin tracking them, before they are significantly affected by ram pressure.

The left panel of Figure \ref{fig1} shows the initial mass of neutral and ionized
gas in each galaxy, with the color representing the stellar mass. Our sample
contains a variety of galaxies in stellar and gas mass. Galaxies
with large $M_{*}$ tend to possess very massive hot halos of ionized gas, whereas
lower-mass galaxies have considerable variation in their neutral and ionized gas content.

The right panel of Figure \ref{fig1} shows the morphologies of the gas particle
distribution separated into neutral and ionized gas, as well as that of the stellar
component of each galaxy. Ellipticity is shown on
the horizontal axis, such that more spherical objects are to the left of the plot,
while triaxiality is shown on the vertical axis, such that oblate objects are to the lower
right of the plot and prolate ones are to the upper right.
Despite the fact that the neutral gas component was not
chosen based on morphology, most of the galaxies have neutral gas that lies
in a flattened (oblate) disk. The stellar component of the majority of the galaxies
is also flattened, with a minor-to-major axis ratio of $<0.5$. The color of the points
representing the stellar and neutral gas components indicates the relative alignment
of the minor axes of the two. We see that in the majority of cases, the 
stars and neutral gas are aligned within $20^{\circ}$.

The majority of galaxies have a prolate distribution of the ionized gas particles bound
to their subhalo. However, as we will see in \S \ref{results_analytic}, 
the density profile of the ionized gas can still be well-approximated as spherical.

\section{Ram Pressure Models} \label{models}

In this section, we first describe the analytic ram pressure models from the literature
that will be used to predict the ram pressure stripping of 
the neutral gas disk and ionized halo, and will be compared to the 
stripping rate from the EAGLE simulation. We then describe
a slightly more complex toy model that we will also compare to EAGLE
in order to help explain the deviations of the analytic models from the simulation.

\subsection{Analytic Ram Pressure Models} \label{analytic}

The criterion for gas to be stripped from a galaxy by ram pressure
was first described in \citet{gunngott}. Gas can be stripped
when the force of the ram pressure exceeds the restoring force
exerted on the gas by the gravitational potential of the galaxy. This
can be expressed as:
\begin{equation}
\label{eq1}
P_{\mathrm{ram}} > \Sigma_{\mathrm{gas}}g_{\mathrm{max}}.
\end{equation}
Here $\Sigma_{\mathrm{gas}}$ is the mass surface density of the galaxy gas integrated
along the ram pressure direction, and $g_{\mathrm{max}}$ is the maximum
restoring acceleration along the direction of the ram pressure.
$P_{\mathrm{ram}}$ is the ram pressure, which is given by
\begin{equation}
P_{\mathrm{ram}} = \rho_{\mathrm{ICM}} v_{\mathrm{rel}}^{2},
\end{equation}
where $\rho_{\mathrm{ICM}}$ is the density of the intracluster medium (ICM) through
which the galaxy is passing, and $v_{\mathrm{rel}}$ is the relative velocity between the
galaxy and the ICM.

We compute $\rho_{\mathrm{ICM}}$ and $v_{\mathrm{rel}}$ from the local ICM
properties around each galaxy at each simulation output. The relevant ICM particles 
are taken to be those that pass through a sphere containing
all the gas bound to each galaxy, or through a sphere with 30 kpc radius if larger 
\footnote{The median number of ICM gas particles used to compute the ram pressure is 
1081, with interquartile range 302 to 3456. $3.8\%$ of outputs
have ram pressure estimates based on fewer than 50 particles;
this is most commonly caused by very low ICM velocities relative to the galaxy, 
such that few ICM particles pass through the sphere containing the galaxy.}.
ICM particles with density above $5\times 10^{13} \msun/$Mpc$^{3}$ and temperature less
than $10^{7}$ K are excluded from consideration to avoid selecting cold, dense substructures within the cluster. 
The relative velocity is $v_{\mathrm{rel}} = 
v_{\mathrm{ICM}} - v_{\mathrm{CoM}}$, where $v_{\mathrm{CoM}}$ is the center-of-mass velocity
of the galaxy subhalo, and $v_{\mathrm{ICM}}$ is the ICM velocity. 
$v_{\mathrm{CoM}}$ is computed from the subhalo particles,
but the ICM particles have a range of velocities. To obtain a single direction
and magnitude for $v_{\mathrm{rel}}$, we take its direction $\hat{v}_{\mathrm{rel}}$ to be the one that
maximizes the value of $\sum m_{i} \vec{v}_{i} \cdot \hat{v}_{\mathrm{rel}}/\sum m_{i}$, where $m_{i}$ and 
$\vec{v}_{i}$ are the masses and velocities of the ICM particles selected as described above. The
summed quantity is taken to be the magnitude of the relative velocity.
The median density of ICM particles with relative velocity direction 
within 60$^{\circ}$ of $\hat{v}_{\mathrm{rel}}$ is taken to be the value of $\rho_{\mathrm{ICM}}$.
The ICM velocity is assumed to be unidirectional and constant between simulation outputs.
We find that our ram pressure magnitude estimates are generally similar to those computed using
the prescription given in \citet{marasco}, even with small numbers of ICM particles
passing through the galaxy.

\begin{figure*}
    \centering
    \includegraphics[width=\linewidth]{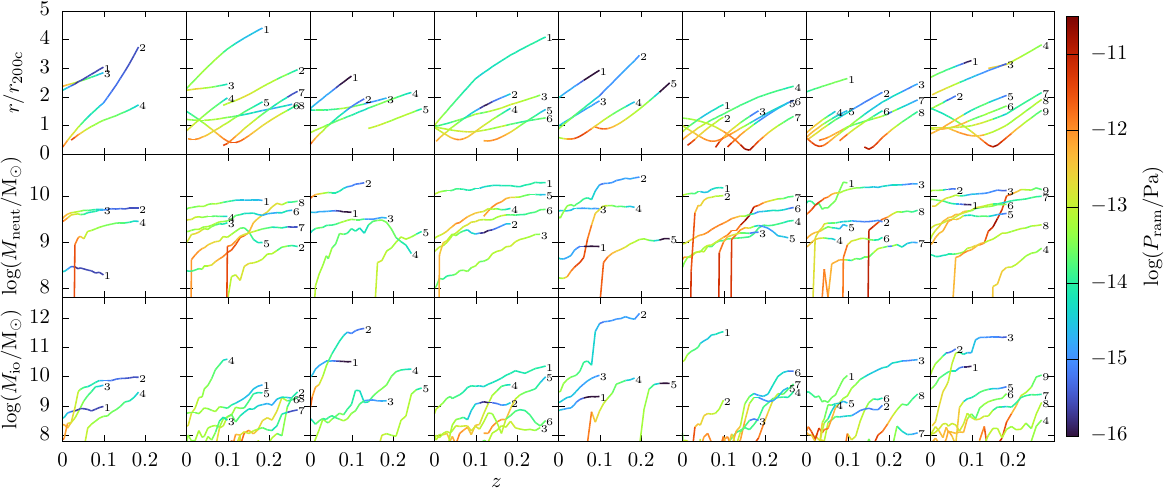}
    \caption{\label{fig2} The evolution of a subset of the galaxies in our 
    sample within the eight $z = 0$ EAGLE clusters
    ($M_{200c} > 10^{14} \msun$). Each column represents one of the 8 clusters, in order of 
    increasing $M_{200c}$ from left to right. The color scale in each panel shows the ram pressure 
    computed from the local ICM as described in \S \ref{analytic}. For each cluster, galaxies
    are labeled with numbers to allow for identification between rows.
    \textit{Top row:} The trajectory of each galaxy as a function of redshift, expressed as the 3D distance from the 
    center of the cluster in terms of its $r_{200c}$.
\textit{Middle row:} The neutral gas mass of each galaxy as a function of redshift.
\textit{Bottom row:} The ionized gas mass of each galaxy as a function of redshift.}
\end{figure*}

The ram pressure computed in this manner is presented as the color scale in Figure \ref{fig2}, 
in which we show the trajectories of a subset of galaxies in our 
sample as they fall into their host clusters, 
as well as the evolution of their neutral and ionized gas masses.
We see in the top row of panels that the ram pressure typically 
increases with decreasing distance to the center of 
the cluster, as expected. The second row of panels shows the evolution 
of the neutral gas mass of each galaxy, indicating that typically 
large ram pressures ($P_{\mathrm{ram}} \gtrsim 10^{-12.5}$ Pa) are required to efficiently strip neutral gas from 
the disk. Such conditions in most cases occur significantly within $r_{200c}$, during the galaxy's pericenter
passage. By contrast, the bottom row of panels shows that the ionized halo can be stripped by lower ram pressures,
which occur even outside of $r_{200c}$.

For the right-hand term in Eqn. \ref{eq1}, we will adopt different analytic approximations
for the neutral gas disk and the quasi-spherical ionized gas halo: the former
from \citet{gunngott}, and the latter from \citet{mccarthy}.

In \citet{gunngott}, an analytic approximation for Eqn. \ref{eq1} is given
for the case of a perfectly flat stellar and gas disk whose motion relative
to the ICM is perpendicular to the disk direction (face-on):
\begin{equation}
\label{gmaxgg}
g_{\mathrm{max}} = 2\pi G \Sigma_{*},
\end{equation}
where $\Sigma_{*}$ is the stellar disk mass surface density. Thus the analytic stripping
criterion for the neutral gas disk would be:
\begin{equation}
\label{eqgg}
P_{\mathrm{ram}} > 2\pi G \Sigma_{*} \Sigma_{\mathrm{gas}}.
\end{equation}

We implement this model by computing the stellar and gas surface mass densities
in the direction perpendicular to the plane of the stellar disk (which, as shown
in Figure \ref{fig1}, is typically well-aligned with the neutral gas disk).
We obtain these surface densities by smoothing the particles 
with a 2D Wendland C2 kernel \citep{Wendland} with 58 neighbors 
\footnote{This kernel is akin to that used for SPH smoothing in EAGLE, but 2D rather than 3D.}.
Since the gas density in the simulation is based on smoothing 
over all gas particles regardless of their
subgroup membership, for the gas mass surface density we also smooth over the nearest
20 kpc of ICM particles. To verify that the results are robust to this choice, we also computed them
using the gas surface density from only the gas particles in the galaxy. 
By construction, this underpredicts the simulation gas surface density; 
however, while the resulting stripping timescales 
are shorter, our results remain qualitatively similar.

For each simulation output, we compare the value of $P_{\mathrm{ram}}$, computed
as described above, to the value of $\Sigma_{\mathrm{gas}}g_{\mathrm{max}}$ for
each gas particle. Once $P_{\mathrm{ram}}$ exceeds $\Sigma_{\mathrm{gas}}g_{\mathrm{max}}$,
we consider the particle to have been stripped in the analytic model.

An analytic approximation for a spherical halo, such as that in which
the hot ionized gas typically resides, is given in \citet{mccarthy}. Assuming
isothermal gas and dark matter halo profiles, they find:
\begin{equation}
\label{sigma_gmax_halo}
g_{\mathrm{max}} = \frac{GM(<r)}{2r^{2}} \mathrm{\: and \:}
\Sigma_{\mathrm{gas}} = \pi r \rho_{\mathrm{gas}}(r),
\end{equation}
where $M(<r)$ is the total mass of the subhalo within radius $r$, and
$\rho_{\mathrm{gas}}(r)$ is the gas density at $r$. This gives the
stripping criterion as:
\begin{equation}
P_{\mathrm{ram}} > \frac{\pi}{2} \frac{GM(<r)\rho(r)}{r}.
\end{equation}
However, for more general power-law profile models for the gas and dark matter halos,
the resulting equation is:
\begin{equation}
\label{eqmc}
P_{\mathrm{ram}} > \alpha \frac{GM(<r)\rho(r)}{r},
\end{equation}
where $\alpha$ depends on the choice of profiles. \citet{mccarthy} find 
$\alpha = 2$ to be the best fit to the results of their hydrodynamical simulations.

In our implementation of this model, we use a power-law fit to the
density of the ionized gas particles for $\rho(r)$. We note that the density
of each particle in EAGLE is computed using kernel smoothing over \textit{all}
nearby gas particles, including those not part of the ionized halo.

We also attempted to use a power law for the total subhalo mass profile $M(<r)$,
but found that it was poorly fit by any single power law, instead typically being well-described
by an NFW profile \citep{nfw}. We thus use the true subhalo mass profile
measured around the most bound particle when computing Eqn. \ref{eqmc},
which includes the mass from all particle types (dark matter, stars, and gas).

Additionally, we tried using the radially smoothed density of the ionized
gas for $\rho(r)$ rather than a power law fit. 
This produced qualitatively similar results to those presented in \S\ref{results_cumulative}.

\subsection{Ram Pressure Toy Model}\label{toymodel}

The analytic models described in the previous subsection rely on a number
of idealized assumptions, specifically regarding the geometry of the galactic gas 
and gravitational potential well, as well as the orientation of the galaxy relative to the ram pressure direction. 
In this subsection, we describe a toy model that treats these in a more realistic fashion while nevertheless
being more simplified than a full hydrodynamical simulation. We use this model to investigate
whether adding to the complexity of analytic models can improve their agreement
with hydrodynamical simulations. Additionally, we use it to systematically analyze
the physical reasons why the analytic model differs from the simulation.

The toy model is more complex than the analytic models 
in that the gravitational potential distribution
of the subhalo and the orientation of the galaxy
relative to the ram pressure direction are taken from the simulation,
and the orbits of gas particles are integrated rather than the direction
of stripping being assumed to be vertical from the initial position. The 
inclination of a galaxy relative to the ram pressure is known to make a difference
to the stripping rate \citep{roetilt}, and can cause changes to the gas morphology,
such unwinding the spiral arms of galaxies that are being stripped edge-on \citep{bellhouse}.

The toy model is still simplified relative to hydrodynamical simulations
in a number of respects. 
The gas particles are treated as test particles within the subhalo
gravitational potential,
and hydrodynamical interactions between the gas particles are neglected, as are all subgrid
physics effects. Because the dark matter is the largest contributor to the gravitational
potential of the subhalo and is unaffected by ram pressure, we find that the evolution
of the subhalo potential due to change in the distribution of gas 
between outputs spaced by $\approx 120$ Myr is generally small. However,
as we will describe in \S\ref{results_toy}, hydrodynamical effects and subgrid physics have 
significant influence on the ram pressure stripping process in EAGLE.

The toy model also distributes the force of the ram pressure over the galactic gas in a 
simplified fashion. As described in \S \ref{analytic}, 
the ram pressure is computed from the local ICM properties as 
$P_{\mathrm{ram}} = \rho_{\mathrm{ICM}} v_{\mathrm{rel}}^{2}$. Unlike
for the analytic models, in the toy model 
$v_{\mathrm{rel}} = v_{\mathrm{ICM}} - v_{\mathrm{gas}}$, where
$v_{\mathrm{gas}}$ is the velocity of each individual gas particle in the galaxy.
The acceleration of each gas particle is then estimated 
as $a_{\mathrm{ram}} = P_{\mathrm{ram}}/\Sigma_{\mathrm{gas}}$, where 
$\Sigma_{\mathrm{gas}}$ is the gas surface density integrated along the ram pressure velocity 
direction, computed as described in \S\ref{analytic}. This 
formulation for the acceleration due to ram pressure is oversimplified, as it
does not take into account the fact that gas at the front of the galaxy ``shields''
the gas behind it from ram pressure \citep{tilted}. Using a value
of $\rho_{\mathrm{ICM}} v_{\mathrm{rel}}^{2}$ for the ram pressure magnitude
also relies on the assumption that the ICM wind deposits its full momentum into 
the gas of the oncoming galaxy. As we will discuss is \S\ref{confinement},
this assumption is inaccurate.

We recompute $\Sigma_{\mathrm{gas}}$ as we evolve the motions of the gas particles in the toy model, and estimate
the acceleration due to ram pressure using the instantaneous value of $\Sigma_{\mathrm{gas}}$.
By contrast, we update the gravitational potential of the subhalo to that in the simulation only at the
beginning of each integration, and also change the ram pressure magnitude and direction
based on the local ICM properties only at this time.

The gravitational potential within each subhalo is interpolated from the outputs of \textsc{subfind}.
This interpolation does not extend outside the
boundaries of the subhalo identified by \textsc{subfind}, which uses saddle points in
the density distribution to define the physical extent of subhalos. When particles 
exit this region, we assign them a potential of 0, such that they are always unbound. 

We make two different types of comparisons using the toy integration. For the first, we 
select all the gas particles bound to the galaxy in the initial simulation output
at which we identify it (before it enters the cluster FoF group), and integrate
the motion of these particles under the assumptions above for the entire
time that we track the galaxy in the simulation. That is, we use the positions
and velocities that we obtain for the gas particles at the end of each integration
to be the initial conditions for the next integration, but 
but we update the subhalo potential and ram pressure magnitude and direction
to be those computed for the new simulation output.
We ignore any new gas particles that are accreted to the galaxy since the time
at which we begin tracking it, although as described in \S\ref{galsamp}, we have selected
galaxies that accrete very little gas.
We will use these results to compare the stripped mass 
fraction predicted by the analytic models, toy model, and simulation,
and to identify the main reasons why the analytic models differ from the simulation.

For the second comparison, we will perform the toy model integration separately between individual
simulation outputs. We will then compare each toy model result to the subsequent output.
This approach should result in less divergence from the simulation results, as the toy model
integration is run for only a short time. We will perform the second
comparison only between the toy model and the simulation, in order to present subgrid
and hydrodynamical effects that have a significant impact on the outcome of ram pressure stripping.

We are interested in comparing the fraction of gas mass stripped in the toy model,
analytic models, and simulation. However, this depends on the definition used
for stripped gas. In the analytic models described in \S \ref{analytic},
gas is assumed to be stripped when the force of the ram pressure exceeds the gravitational
restoring force on the gas. As discussed in, for example,
\citet{rp2018}, this is not identical to the gravitational unbinding of the gas, 
which is when the gas velocity exceeds the escape speed of the subhalo. 
Gas for which the maximum restoring force has been exceeded
is generally not yet energetically unbound. Additionally, gas that experiences a 
rapid impulse of ram pressure can achieve escape velocity and become energetically
unbound despite the ram pressure 
not exceeding the maximum gravitational restoring force.

EAGLE does not output any data related to the acceleration experienced by particles, 
and thus we do not have a record of whether every gas particle exceeded its maximum restoring
force to compare directly to the analytic models. For the simulation,
we compute the sum of the kinetic and potential energy of each gas particle 
and designate the particle as unbound if this value is larger than zero\footnote{In EAGLE, thermal
energy also contributes to whether a gas particle is considered energetically unbound or not, but here
we are only interested in a kinematic comparison, and we thus ignore this term. 
For neutral gas, which is cold, the median ratio of the thermal energy to the 
sum of the kinetic and potential energies is less than $1\%$. 
For the hotter ionized gas, the median contribution is $21\%$ for particles in galaxies at 
low ram pressures ($< 10^{-15}$ Pa) and decreases to $4\%$ for particles 
in galaxies at high ram pressures ($> 10^{-12}$ Pa).}. Particles
that exit the region of the \textsc{subfind} subhalo defined by saddle points in the density
are no longer assigned to that subhalo, and are assigned a potential energy of 0.

In the toy model, we compute both the energy of each gas particle 
and whether it exceeded the maximum restoring force it encountered within
its last vertical oscillation, and compare these to the simulation and analytic models,
respectively.

In \S\ref{results_analytic} we will examine the analytic model assumptions
for the initial conditions of the subhalo potential and gas distribution,
and determine the model parameters that best agree with the simulation, which
we will then use throughout the rest of \S\ref{results}. A summary
of the different models described in this section --- analytic, toy,
and simulation --- can be seen in Table \ref{table:results2}, which presents
the outputs of the models for each gas particle and how they are obtained.
This table presumes the analytic model parameters from \S\ref{results_analytic}, which 
will be explained immediately below.

\section{Comparison of Model Results} \label{results}

Here we compare the predictions of the analytic and toy models 
described in \S \ref{models}, and those of the EAGLE reference simulation described in \S \ref{eagle}.

\subsection{Evaluation of Analytic Model Initial Assumptions} \label{results_analytic}

\begin{figure*}
    \centering
    \includegraphics[width=0.9\linewidth]{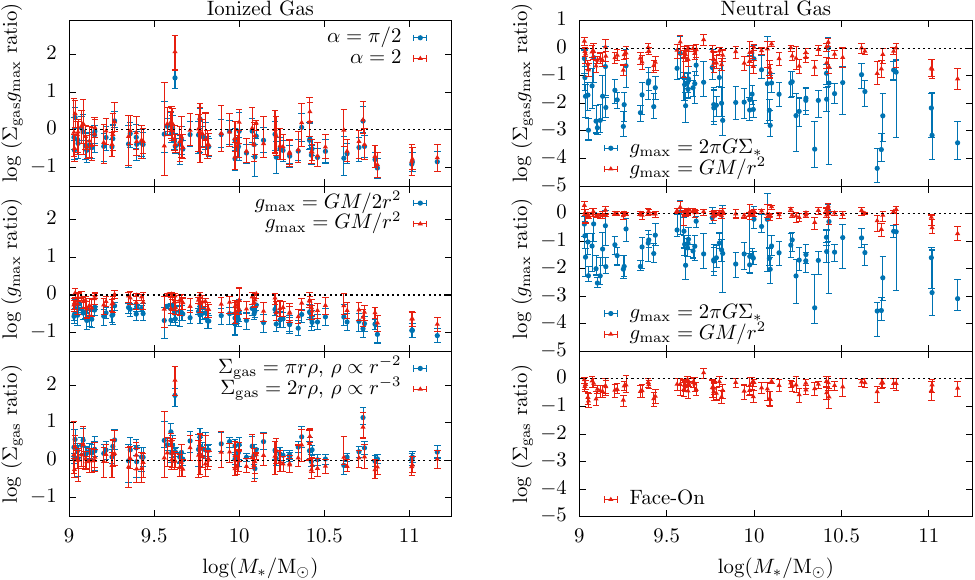}
    \caption{\label{fig3} Ratio between galaxy parameters assumed by the analytic models 
    ($\S$\ref{analytic}) and the initial parameters from the simulation/toy model 
    (see \S\ref{results_analytic} for explanation), as a function of initial galaxy stellar mass. 
    Parameters for the ionized gas halo are presented on the left and those for the neutral gas 
    disk on the right. From bottom to top, the parameters shown are gas surface density $\Sigma_{\mathrm{gas}}$,
    maximum gravitational restoring acceleration $g_{\mathrm{max}}$, 
and $\Sigma_{\mathrm{gas}} g_{\mathrm{max}}$, which must be exceeded by the ram pressure to strip the gas particle. 
    Each of the 80 galaxies in our sample is represented by
a point with error bars, indicating the median and 
    quartile values of the ratio of analytic to simulation/toy parameters for all the 
    gas particles initially bound to the galaxy. 
    Different point colors show different assumptions 
    for the analytic models presented in \S\ref{analytic}, as explained in \S\ref{results_analytic}.}
\end{figure*}

As described is \S \ref{analytic}, we employ two analytic models 
from the literature to 
predict the total (cumulative) gas mass stripped from the 
neutral gas disk and ionized hot gas halo of galaxies. 
These models make assumptions about the geometry of these 
gaseous components, as well as about the
gravitational potential in which the gas resides.

In Figure \ref{fig3}, we present the ratio of the 
initial parameters assumed in the analytic models to 
the initial parameters from the simulation/toy model, as a function
of the initial stellar mass of each galaxy. 
The left panels show the comparison for the 
ionized gas, which in the analytic models is assumed 
to be spherically distributed and also reside in 
a subhalo with spherically symmetric mass distribution. 
The right panels show the comparison for the neutral gas,
which is assumed to be in a flat disk
perpendicular to the direction of the ram pressure, with the
restoring force produced by a stellar disk in the same plane.
This is described in detail in \S\ref{analytic}.

The top panels of Figure \ref{fig3} show the
log of the ratio of $\Sigma_{\mathrm{gas}} g_{\mathrm{max}}$ 
assumed in the analytic model, to that in the 
toy model at the initial time. 
Specifically, $\Sigma_{\mathrm{gas}}$ is the gas mass surface 
density at the location of each gas particle,
while $g_{\mathrm{max}}$ is the maximum gravitational restoring acceleration
opposing the ram pressure that is encountered by each gas particle. 
The center and lower panels of Figure \ref{fig3} directly compare $g_{\mathrm{max}}$
and $\Sigma_{\mathrm{gas}}$ from the analytic models
to their values from the simulation/toy model.

For the toy model, $\Sigma_{\mathrm{gas}}$ is
measured directly from the gas distribution in the simulation, 
integrated in the direction of the ram 
pressure in the initial simulation output. $g_{\mathrm{max}}$ is the maximum restoring 
acceleration in the first vertical orbit of each gas particle 
(which can be multiple outputs long); this value is not reported
by the EAGLE simulation and is thus taken from the toy model integration.
The analytic models for $\Sigma_{\mathrm{gas}}$ and $g_{\mathrm{max}}$
of the neutral gas disk and ionized gas halo are presented in \S \ref{analytic}.

For the ionized halo analytic model, $\Sigma_{\mathrm{gas}} g_{\mathrm{max}}$ has the form described in Eqn.
\ref{eqmc}. The top left panel of Figure \ref{fig3} shows 
the distribution, in terms of median and quartiles,
of the log of the ratio of $\Sigma_{\mathrm{gas}} g_{\mathrm{max}}$ for the ionized
gas particles in each of the 80 galaxies in our sample. If the toy model and analytic model
were always in agreement, the value would be zero.
If $\alpha$ in Eqn. \ref{eqmc} is taken to be $\pi/2$, as it would be if both the gas and total
matter distribution of the subhalo had an isothermal density profile, 
$\log (\Sigma_{\mathrm{gas}} g_{\mathrm{max}})$ for the analytic model 
differs from that for the simulation/toy model by $-0.32 \pm 0.04$ dex (blue points). 
Alternatively, the red points show $\alpha = 2$, which \citet{mccarthy} 
found was in better agreement with their hydrodynamical simulations, and which we
find to perform slightly better than $\alpha = \pi/2$, 
being on average different from the toy model by $-0.20 \pm 0.05$ dex.

Greater disagreement can be seen when comparing $\Sigma_{\mathrm{gas}}$ and $g_{\mathrm{max}}$ individually
between the analytic and toy models. This is presented for the ionized gas in the center left
and lower left panels of Figure \ref{fig3}. In the center left panel,
we see as the blue points $g_{\mathrm{max}} = GM/2r^{2}$, which is the form for 
an isothermal dark matter halo. Note, however, that we have used the actual
subhalo mass profile from the simulation for $M(<r)$, as described in \S\ref{analytic},
since the mass profile is described by an NFW form rather than any single power law. Here
we see that this form gives a value for $g_{\mathrm{max}}$ on average $-0.56\pm0.02$ dex
away from the toy model value. The maximum possible finite
value for $g_{\mathrm{max}}$ from a power-law model is $g_{\mathrm{max}} = GM/r^{2}$, corresponding to
to $M \propto r^{2}$. This is plotted as the red points, again taking the simulation
mass profile for $M(<r)$ rather than a power law. We see that it is a better fit to the toy model value
of $g_{\mathrm{max}}$, but still differs on average by $-0.26\pm0.02$ dex, despite
giving the highest possible values for a power-law coefficient. This is likely because,
as we note, the subhalo mass profile is not well fit by a single power law.

The ratios in Figure \ref{fig3} are plotted as a function of the initial stellar masses of the galaxies. 
We see that for $g_{\mathrm{max}}$ of the ionized gas, the discrepancy between the analytic model
and the toy model increases with increasing galaxy mass. Galaxies with the highest
stellar masses initially possess very massive ionized halos of gas (see Figure \ref{fig1})
that cause the total subhalo mass profile to deviate from an NFW.
All other properties in Figure \ref{fig3} have little correlation with the galaxy stellar mass.

The lower left panel of Figure \ref{fig3} compares $\Sigma_{\mathrm{gas}}$ between the analytic and toy
models for the ionized halo, with blue points showing the results for 
an $r^{-2}$ (isothermal) fit to the ionized gas density profile, and the red points showing
the results for an $r^{-3}$ fit. We see that the latter fit is better: the mean deviation
from $\Sigma_{\mathrm{gas}}$ in the simulation is only $0.07\pm0.03$ dex, whereas for the isothermal fit
it is $0.26\pm0.03$ dex. We note that $r^{-3}$ would correspond to the density profile 
of the ionized halo following that of the dark matter for the outer part of an NFW profile.

While $g_{\mathrm{max}}$ and $\Sigma_{\mathrm{gas}}$ are not individually well fit by the assumption 
of isothermal profiles for both the subhalo and gas mass profiles, it is notable that
in combination they nevertheless produce reasonable values for $\Sigma_{\mathrm{gas}} g_{\mathrm{max}}$
for the ionized halo. Nevertheless, we will take $g_{\mathrm{max}} = GM/r^{2}$ and 
$\rho_{\mathrm{gas}} \propto r^{-3}$ when
comparing the analytic model to the toy model and simulation in \S\ref{results_cumulative},
as these parameters produce slightly better results and are in agreement with the findings
of \citet{mccarthy} regarding the value of $\alpha$.

The right panels of Figure \ref{fig3} similarly compare the analytic model to the initial
conditions of the toy model, but for the neutral gas disk,
the analytic model for which is given by Eqn. \ref{eqgg}.
The blue points in the top panels show the ratio of $\Sigma_{\mathrm{gas}} g_{\mathrm{max}}$,
where we see that the formulation from \citet{gunngott} differs from
the expected value from the toy model by an average of $-1.70\pm0.09$ dex --- i.e., the ram
pressure required to strip the gas is underestimated by a factor of $\sim 50$.
We also show the ratio using $g_{\mathrm{max}} = GM/r^{2}$, the
maximum restoring acceleration we have chosen for the ionized gas, as red points.
Since the neutral gas disk and ionized gas halo reside in the same potential,
they should experience the same maximum restoring acceleration, and this 
approximation to $\Sigma_{\mathrm{gas}} g_{\mathrm{max}}$ is more similar
to the simulation/toy model, differing by only $-0.34\pm0.03$ dex.

The right center panels again show the ratio of $g_{\mathrm{max}}$ for
the analytic model to that from the toy model, now for the neutral disk.
This underscores the fact that Eqn. \ref{gmaxgg} underestimates the gravitational
force experienced by the gas particles: the value of $g_{\mathrm{max}}$
differs from that of the toy model by an average of $1.36\pm0.08$ dex (blue points). By contrast,
assuming $g_{\mathrm{max}} = GM/r^{2}$, where $M$ is the measured mass profile
of the subhalo, the deviation from $g_{\mathrm{max}}$ in the toy model is 
a negligible $-0.01\pm0.02$ dex (red points).
This agreement is even better than for the ionized halo, likely because the neutral gas disk 
lies at the center of the subhalo where the NFW profile
is similar to a power law with $M \propto r^{2}$. We conclude that
taking into account the mass contribution from dark matter rather than only
stars strongly affects the predicted ram pressure stripping of the neutral gas disk,
and adopt $g_{\mathrm{max}} = GM/r^{2}$ for the neutral disk for the remainder of
\S\ref{results}.

Finally, in the bottom right panel, we see the ratio of $\Sigma_{\mathrm{gas}}$ for the 
neutral gas disk in the face-on direction to
that using the true initial ram pressure direction. As expected, assuming
that the galaxy is falling face-on into the ICM minimizes the value of $\Sigma_{\mathrm{gas}}$.
However, the difference is less than one magnitude, which
is smaller than that from not taking into account the effect of dark matter
on the gravitational restoring force.

We conclude from Figure \ref{fig3} that it seems possible to achieve a good fit
to the initial conditions for both the ionized gas halo
and neutral gas disk using relatively simple analytic models for 
the gas surface density and maximum restoring acceleration. Nevertheless, we see across all
the panels that there is variation in the level of agreement between the analytic
and toy models within the galaxy sample, presumably due to the differing
gas and dark matter geometries in real galaxies relative to an idealized approximation.

For the remainder of this section, when comparing the analytic formulations to the toy
model and simulation, we use the better-fitting analytic models in Figure \ref{fig3}.
That is, we use $\alpha = 2$ for the ionized halo analytic model (defined in Eqn. \ref{eqmc}) and 
$g_{\mathrm{max}} = GM/r^{2}$ for the neutral gas disk (rather than Eqn. \ref{gmaxgg}) 
in the following comparisons. These choices, as well as those made for the toy model, are summarized in Table \ref{table:results2}.

\begin{deluxetable*}{c|c||c|c|c|c|}
\tabletypesize{\footnotesize}
\tablecolumns{6}
\tablewidth{0pt}
\tablecaption{\label{table:results2}Summary of Computed Quantities for $i^{\mathrm{th}}$ Gas Particle at Each Simulation Output Time}
\tablehead{ & & $\Sigma_{\mathrm{gas}, i}$ & $g_{\mathrm{max}, i}$ & $\Phi_{i}$ & $\vec{r}_{i}$, $\vec{v}_{i}$ }
\startdata
& Disk & 
$\int\rho_{\mathrm{gas}}(r)$ along disk minor 
& 
& - & -  \\
Analytic & & {axis at \textbf{initial} time} & $GM(<r_{i})/r_{i}^{2}$ at & & \\
\cline{2-3}
Model & Halo & $2r_{i}\rho_{\mathrm{gas}}(r_{i})$ with $\rho_{\mathrm{gas}} \propto r_{i}^{-3}$ & \textbf{initial} time & & \\
& & at \textbf{initial} time & & & \\
\hline
 \multicolumn{2}{c||}{Toy Model} & 
 $\int\rho_{\mathrm{gas}}(r)$ along  $\hat{v}_{\mathrm{rel}}$ direction
 & $\max(-\vec{g}(\vec{r}_{i}) \cdot \hat{v}_{\mathrm{rel}})$ during & $\Phi(\vec{r}_{i})$; $\Phi(\vec{r})$ from each & \textit{\textbf{cumulative}} \\
  \multicolumn{2}{c||}{(Cumulative)} & at \textbf{each} time, starting with & each vertical orbit; $g=-\nabla\Phi$; & simulation output, $\vec{r}_{i}$ from & prediction \\
  \multicolumn{2}{c||}{} & \textit{\textbf{predicted}} gas positions & $\vec{r}_{i}$ from cumulative prediction & \textit{\textbf{cumulative}} prediction & \\
  \hline
 \multicolumn{2}{c||}{Toy Model} & 
  $\int\rho_{\mathrm{gas}}(r)$ along  $\hat{v}_{\mathrm{rel}}$ direction
 & - & $\Phi(\vec{r}_{i})$; $\Phi(\vec{r})$ from each & \textit{\textbf{single-step}}\\
  \multicolumn{2}{c||}{(Single-Step)} & at \textbf{each} time, starting with & & simulation output, $\vec{r}_{i}$ from & prediction \\
  \multicolumn{2}{c||}{} & \textit{\textbf{simulation}} gas positions & & \textit{\textbf{single-step}} prediction & \\
  \hline
   \multicolumn{2}{c||}{EAGLE} & - & - & direct output & direct output \\
   \multicolumn{2}{c||}{Simulation} &  &  & from simulation & from simulation\\
\enddata
\tablecomments{The quantities obtained from
the models described in \S\ref{results_cumulative} and \S\ref{results_toy}, as well
as the EAGLE simulation, at each simulation output time. The relevant quantities for each gas particle are: 
the local gas surface density $\Sigma_{\mathrm{gas}}$, 
the maximum gravitational restoring acceleration experienced by the particle $g_{\mathrm{max}}$, 
the gravitational potential energy per unit mass $\Phi$, 
the position $\vec{r}$, and the velocity $\vec{v}$. 
Other quantities mentioned in the table above include $\rho_{\mathrm{gas}}$,
the 3D density of the gas, and $\hat{v}_{\mathrm{rel}}$, the unit vector in the direction of the relative velocity
between the ICM and the galaxy (i.e. the direction of the ram pressure force). 
The criterion for ram pressure stripping from \citet{gunngott} is 
$P_{\mathrm{ram}} > \Sigma_{\mathrm{gas}}g_{\mathrm{max}}$.
The criterion for a particle being unbound (exceeding the escape speed) is $E/m = \Phi + v^{2}/2 \geq 0$.
The results of the analytic model, cumulative toy model, and simulation are compared in \S\ref{results_cumulative}.
The results of the single-step toy model and simulation are compared in \S\ref{results_toy}.}
\end{deluxetable*}

\begin{figure}
    \centering
    \includegraphics[width=\linewidth]{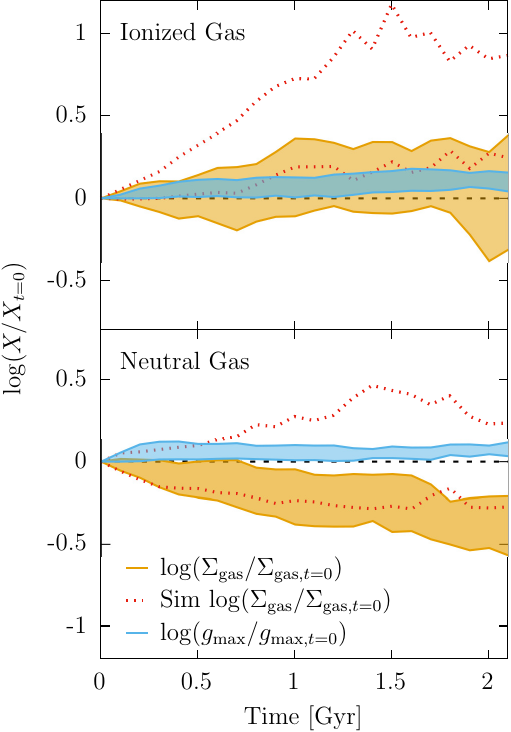}
    \caption{\label{fig4} The evolution of the gas surface density $\Sigma_{\mathrm{gas}}$ and 
    the maximum gravitational restoring acceleration $g_{\mathrm{max}}$ for the gas particles
    in our sample of 80 galaxies. The top panel shows the
    evolution of these values for the ionized gas, and the lower panel for 
    the neutral gas. At each time, we compute the ratio of
    $\Sigma_{\mathrm{gas}}$ and $g_{\mathrm{max}}$ to their initial values
    for each of the ionized or neutral gas particles that are still bound to each galaxy, and then take
    the median value for each galaxy. We then show the quartiles encompassing these
    median values as the shaded region in each panel. The light blue region represents the range of the
    evolution of $g_{\mathrm{max}}$ in the toy model. The orange region
    is the evolution of $\Sigma_{\mathrm{gas}}$ in the toy model. The red dashed
    lines represent the evolution of $\Sigma_{\mathrm{gas}}$ in the simulation,
    computed by taking the true (rather than predicted from the toy model)
    $\Sigma_{\mathrm{gas}}$ in the direction of the ram pressure at each time,
    for the particles that are still bound in the simulation (rather
    than the toy model).}
\end{figure}

\subsection{Evolution of the Surface Density and Maximum Restoring Acceleration} \label{results_evol}

We have just seen that the analytic models described in \ref{results_analytic}
are capable of reproducing the initial values of the gas mass surface density $\Sigma_{\mathrm{gas}}$ (from
the simulation) and the maximum gravitational restoring acceleration 
$g_{\mathrm{max}}$ (inferred from the toy model) within $\sim 0.3$ dex,
given the correct choice of parameters. These analytic models
make the simplifying assumption that
the direction of the ram pressure is constant throughout
the stripping process, and that the gas particles do not
change their distribution in the perpendicular plane over that time.
As a result, $\Sigma_{\mathrm{gas}}$ and $g_{\mathrm{max}}$ are assumed to be constant values
in these models.

However, both the effective $\Sigma_{\mathrm{gas}}$ and $g_{\mathrm{max}}$ can evolve
as a real galaxy is stripped by ram pressure. \citet{rp2018}
note that $g_{\mathrm{max}}$ can change because the force of ram pressure
can alter the orbits of gas particles. Additionally,
the gravitational potential experienced by the gas changes
over time as the gas is stripped. The gas surface density can also change
due to the fact that the ram pressure changes direction
over the orbit of the galaxy in the cluster, causing
the gas distribution perpendicular to the ram pressure
direction to vary. Because the toy model evolves
the orbits of the particles under the ram pressure force,
and also updates the ram pressure direction
and gravitational potential of the subhalo at each simulation output, it should account
for all these effects, albeit imperfectly.

In Figure \ref{fig4}, we show the range of evolution of $\Sigma_{\mathrm{gas}}$
and $g_{\mathrm{max}}$ predicted by the toy model.
At each simulation output, we take the gas particles
in each galaxy that the toy model predicts are still gravitationally
bound, and compute for each particle
the value of $\Sigma_{\mathrm{gas}}$, the local gas mass surface density at
the particle's location perpendicular to the direction of the ram pressure
in that output, and $g_{\mathrm{max}}$, the maximum gravitational acceleration
in the direction opposing the ram pressure during the vertical orbit
that the particle is undergoing. We then take the ratio
of these values to the initial values of $g_{\mathrm{max}}$ and $\Sigma_{\mathrm{gas}}$ for
those particles. We compute the median of these values for each galaxy
and then display the interquartile range of these median values
as a function of time in Figure \ref{fig4}. The top and bottom panels
show the results for the ionized and neutral gas, respectively.
From the light blue regions in Figure \ref{fig4}, we see that $g_{\mathrm{max}}$
tends to increase slightly for the gas particles in
the galaxies in our sample, but typically by less than 0.1 dex.
This implies that evolution in $g_{\mathrm{max}}$ experienced by
gas particles, due to either changes in their orbits 
or the evolution of the subhalo gravitational potential, is a minor factor in determining
the ram pressure stripping rate.

The situation is different for the evolution of
$\Sigma_{\mathrm{gas}}$, shown as the orange regions. Here we see that
for the neutral gas disk, the gas mass surface density
seen by the ram pressure tends to decrease over time, by a median
value of $\sim0.4$ dex over 2 Gyr. By contrast, $\Sigma_{\mathrm{gas}}$ of the ionized
gas in most galaxies increases slightly, by a median value of $\sim 0.1$ dex. 
There also appears to be a substantial range in the evolution
of the gas surface density of different galaxies for both
the neutral and ionized gas. We would
expect galaxies with decreasing $\Sigma_{\mathrm{gas}}$ to undergo faster
gas removal and those with increasing $\Sigma_{\mathrm{gas}}$ to be stripped more slowly.

However, we also show in Figure \ref{fig4} the evolution of $\Sigma_{\mathrm{gas}}$
from the distribution of gas measured directly from
the simulation rather than that predicted by the toy model.
This is represented by the red dashed lines.
Here we use the particles that are gravitationally bound in
the simulation at each time, rather than those predicted
to be bound by the toy model, but we find that this
is not the main cause of the differences between the
toy model and simulation. We see that $\Sigma_{\mathrm{gas}}$ is higher
for both the neutral and ionized gas in the simulation
than in the toy model, with the difference between
the two increasing with time. We will argue in \S\ref{results_toy} that this is
the result of two hydrodynamical effects that are present in the EAGLE simulation
but not accounted for in the simple toy model:
cooling of the ionized gas, and an inwards force caused by deflection of the ICM flow.

Because the next subsection (\S\ref{results_cumulative}) 
involves comparing the stripping timescales from the toy
model to those from the analytic models described in \S\ref{analytic},
we will use the evolution of $\Sigma_{\mathrm{gas}}$ from the toy model
to color-code the figures there. However, we note that if we
instead use $\Sigma_{\mathrm{gas}}$ from the simulation, the qualitative 
trends that we see are similar, implying that some of the evolution
in the simulation is via the same processes in the toy model, despite
the additional influence of hydrodynamics.

\begin{figure*}
    \centering
    \includegraphics[width=\linewidth]{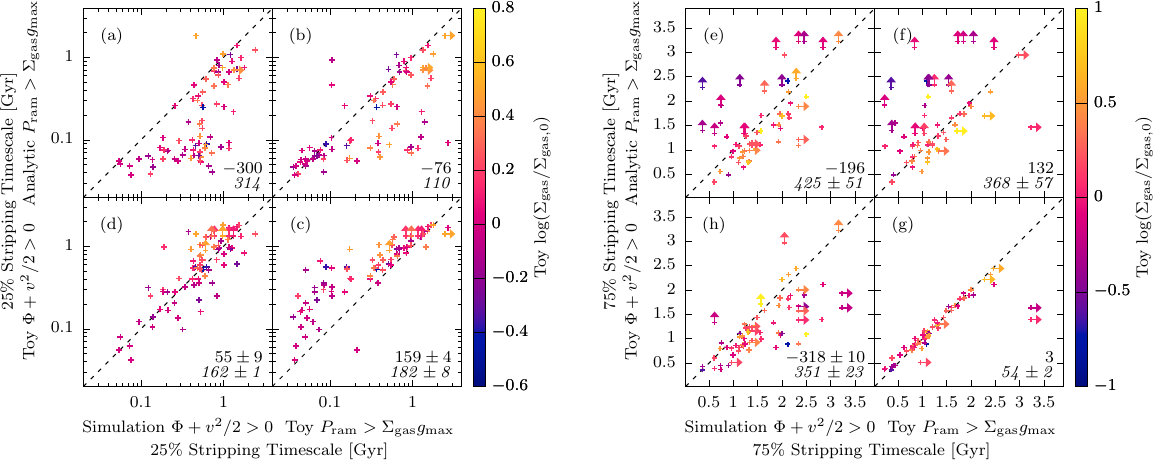}
    \caption{\label{fig5} Gas stripping timescales for the ionized gas in our galaxy sample, 
    given by the simulation as well as by the analytic and toy models described in \S \ref{models}. 
    Panels (a)-(d) on the left show the timescale at which $25\%$ of the ionized gas mass is stripped, whereas 
    (e)-(h) on the right show the timescale at which $90\%$ is stripped. Note that the left panels
    are in log scale as there is a large range in timescales. As our simplified models
    contain no prescription for star formation, we remove those gas particles
    that have transformed into star particles by the next simulation output from consideration when 
    computing stripped fractions. The color scale in the 
    panels represents $\Sigma_{\mathrm{gas}}/\Sigma_{\mathrm{gas}, t=0}$ for each galaxy, 
    i.e. the median ratio of the surface 
    density relative to its value at the initial time in the toy model. 
    This is computed at the time when the toy model reaches the presented unbound fraction 
    ($25\%$ or $90\%$). Where the toy model fails to ever reach the latter fraction, 
    the value of $\Sigma_{\mathrm{gas}}/\Sigma_{\mathrm{gas}, t=0}$ at the final output is shown. Arrows indicate 
    lower limits (where a model or the simulation does not reach the given stripped fraction), 
    but galaxies in which neither the x- or y-axis value 
    achieves the given stripped fraction are excluded from the plot. The numbers in the 
    lower right corner of each panel give the median deviation and median absolute deviation (italics)
    of the y-axis timescale from the x-axis timescale, in Myr. Due to the presence of lower limits,
    the median is sometimes not uniquely determined. In these cases, we give the possible
    range of the median value.
    }
\end{figure*}

\begin{figure*}[t]
    \centering
    \includegraphics[width=\linewidth]{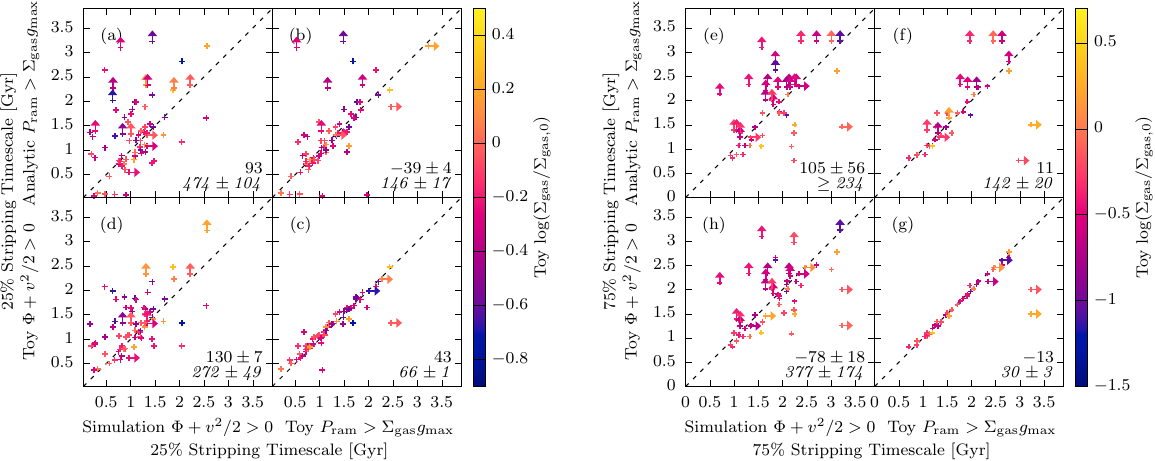}
    \caption{\label{fig6} The same as Figure \ref{fig5}, 
    except all values are shown for the neutral gas,
    and the timescale shown in panels (e)-(h) is for stripping of $75\%$ of the gas, 
    as the neutral disk is not stripped of $90\%$ of its gas for most of the galaxies in
    the toy model and analytic approximations.}
\end{figure*}

\subsection{Cumulative Ram Pressure Stripping Predictions} \label{results_cumulative}

We now compare different measures of cumulative stripping from the analytic model, toy model
integration, and simulation results to determine how much simple analytic models
differ from simulation results and the causes of that difference.

Figures \ref{fig5} and \ref{fig6} present stripping timescales from different models
for the ionized and neutral gas, respectively.
We note that these timescales are measured from
the initial time at which we begin tracking the galaxy, and thus depend on the
distance from the cluster center at which the galaxy enters the cluster FoF group,
as well as the trajectory of the galaxy through the cluster and the cluster properties 
(see Figure \ref{fig2}).
The timescales are linearly interpolated from the stripped fraction at each model/simulation output.

Because our toy and analytic models do not include star formation, gas particles that
become stars by the next simulation output are fully excluded from the gas particle
population when computing the stripped fractions of gas for all of the models. 
Similar results are obtained if excluding at each time all the gas that will 
become stars by the final simulation output at which the
galaxy is considered. For the gas that is initially ionized, typically 
little of it becomes stars by the time that we cease
tracking the galaxy: the range is $0.02\%$ to $17\%$, with median value $4\%$.
However, for most galaxies a substantial fraction
of the initial neutral gas is converted to stars by the time that we cease
tracking them: the range is $1\%$ to $58\%$, with median $28\%$. We note that
these values are the original mass in gas that is converted to stars; star particles in EAGLE lose
mass over time that is added to nearby gas particles, based on a prescription for
stellar mass loss. In the figures presented in this section, we express all stripped mass fractions
as a function of the initial mass of the gas particles.

Panels (a)-(d) of Figure \ref{fig5} present the timescale at which 
$25\%$ of the ionized gas is stripped from each galaxy.
In Panel (a), the horizontal axis is the timescale at which $25\%$ of the gas
is energetically unbound in the simulation, while the vertical axis is the
same timescale for the ram pressure force to exceed the maximum restoring force
($P_{\mathrm{ram}} > \Sigma_{\mathrm{gas}} g_{\mathrm{max}}$)
in the analytic model. We see that the analytic model typically substantially 
underestimates the $25\%$ stripping time. The numbers in the bottom right of the panel give
the median deviation (plain text) and median absolute deviation (italics)
of the vertical axis value from the horizontal axis value, in Myr. Because some values
in each panel are lower limits (arrows) where one of the models did not reach
the given stripped fraction, the mean deviation is not always defined, and therefore sometimes
a range of values is given for the median. This median also excludes those 
galaxies for which both models being compared give lower limits for the stripping timescale (i.e., for
which neither model reached the stripped fraction in question).
The median deviation values in Panel (a) show that the analytic model
typically estimates a $25\%$ stripping timescale for the ionized gas that is
300 Myr shorter than the timescale from the simulation, and the median absolute 
deviation between the timescales is the same (314 Myr).

Panels (b)-(d) use the toy model to elucidate the origin of the difference between the 
analytic model and simulation. Panel (d) again shows the timescale from the simulation, now
versus the $25\%$ unbinding timescale from the toy model. Here we see that that the 
toy model accurately predicts the timescale at which $25\%$ of the ionized gas is gravitationally
unbound from the subhalo: the median deviation from the simulation is only $\approx 55$ Myr, with
a median absolute deviation (scatter) of $\approx162$ Myr. Therefore, it seems that
hydrodynamical and subgrid effects are not the main reason for the difference between
the simulation and the analytic model at this timescale, as they would also be present between
 the simulation and toy model.

Instead, we see in Panel (c) that in the toy model, 
the timescale for the ram pressure force to exceed the maximum restoring force is
shorter than the timescale for the gas to become gravitationally unbound, by
a median value of $\approx 159$ Gyr. This is as expected for gas
exposed to a continuous ram pressure force \citep{rp2018}. This implies that
some of the difference between the analytic model and simulation in Panel (a) is
due to differing definitions of ``stripped'' gas.

Additionally, for the $25\%$ timescale for the ram pressure force to
exceed the maximum restoring force, there are differences between the
analytic model and toy model. This is shown in Panel (b). Here the 
analytic model predicts a stripping timescale slightly shorter than
the toy model, by 76 Myr.

From these panels we find that there is good agreement between the toy model and simulation
predictions for the $25\%$ stripping timescale (Panel d). We also see that the timescale
predicted by the analytic model from \citet{mccarthy} is shorter than that from the simulation (Panel a),
which is partly due to the fact that it takes longer for gas to become energetically unbound 
than it does for it to encounter the maximum restoring gravitational acceleration
during its orbit (Panel c). However, even the analytic timescale to reach the maximum restoring
acceleration is not fully in agreement with that predicted by the toy model (Panel b). 
This is partly due to the fact that the analytic prescription does not include evolution of 
the gas mass surface density, $\Sigma_{\mathrm{gas}}$. Based on Figure \ref{fig4}, 
we expect the evolution of $\Sigma_{\mathrm{gas}}$ to generally 
increase the predicted stripping timescale in the toy model relative to the analytic model, 
and we check this by color-coding each galaxy by its change in $\Sigma_{\mathrm{gas}}$.

The color scale in Panels (a)-(d) represents the same quantity
shown in Figure \ref{fig4} for the toy model --- the median of the ratio of 
each gas particle's $\Sigma_{\mathrm{gas}}$ to its initial value --- at the 
$25\%$ unbound timescale of the toy model for each galaxy. We see in Panel (b) 
that galaxies whose ionized gas tends to increase in surface density
also tend to have longer stripping timescales than the analytic model, which does not include 
evolution of $\Sigma_{\mathrm{gas}}$.

Panels (e) through (h) in Figure \ref{fig5} are similar to panels (a) through (d)
except that they show the timescale at which $90\%$ of the ionized gas is stripped
from each galaxy in the simulation, toy model, and analytic model. Unlike for the $25\%$ stripping
timescale, there does not appear to be a significant difference between the energetic
and force-based criteria for stripping in the toy model, as seen in Panel (g).
There is again some disagreement between the toy model and analytic model, as seen in Panel (f), 
though now the analytic model is typically longer than the toy model (median deviation 132 Myr).
There is also a substantial scatter between the two models 
($\approx 368$ Myr) that has the expected correlation with the evolution
of $\Sigma_{\mathrm{gas}}$ in the toy model (color scale).

Both the analytic and toy models underestimate the $90\%$ stripping 
timescale from the simulation, as seen in Panels (e) and (h). 
The scatter between the analytic model and simulation in Panel (e) correlates with
the evolution in $\Sigma_{\mathrm{gas}}$ in the same way as the scatter
between the analytic and toy models in Panel (f), suggesting that the evolution in $\Sigma_{\mathrm{gas}}$ 
genuinely influences the ram pressure stripping rate.

Figure \ref{fig6} has a similar format to Figure \ref{fig5} but presents 
the results for the neutral gas. The left four panels show the $25\%$ stripping
timescales from the models and simulation, while
the right panels show the same for the $75\%$ stripping timescale. Unlike for 
the ionized gas, the $90\%$ timescale is not shown as the majority of the galaxies do not
reach this level of stripping of their neutral gas.

We see in Panel (c) of Figure \ref{fig6} that the timescale for $25\%$ of the neutral gas particles
to become gravitationally unbound is slightly longer than the same timescale for exceeding the
maximum restoring acceleration, with a median deviation of $43$ Myr.
For the $75\%$ timescale, in Panel (g), the difference is negligible.
In addition to there being little difference between the two definitions for stripping, 
the analytic model is in approximate agreement with the toy model for both the $25\%$ and $75\%$ stripping
timescale (Panels b and f, respectively). 

The majority of the differences between the analytic model
and the simulation, seen in Panels (a) and (e), are also present in the toy model,
seen in Panels (d) and (h). The toy model overestimates the timescale for $25\%$ 
of the gas to become unbound by on average (median) $\approx 130$ Myr,
and there is a significant scatter ($\approx272$ Myr) between the toy model and simulation timescales as well. 
While the median difference in the $75\%$ stripping timescale between the toy
model and the simulation is only $78$ Myr, 
the median absolute deviation between the two is significant, being $377 \pm 174$ Myr.

Additionally, in contrast to the ionized gas, there is no correlation between
the evolution in the gas mass surface density and the difference between the analytic
and toy models. This may be because the evolution in $\Sigma_{\mathrm{gas}}$ is insufficient to significantly
change the neutral gas disk stripping timescales in the toy model, as the
toy model is nearly always in good agreement with the analytic model (Panels b and f).

\begin{figure*}[t]
    \centering
    \includegraphics[width=\linewidth]{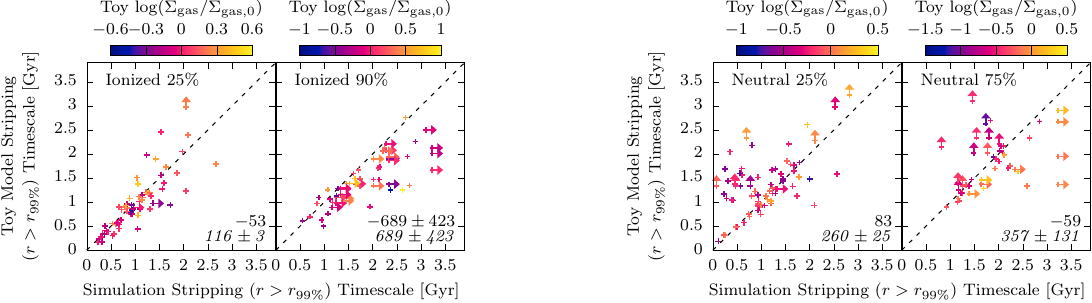}
    \caption{\label{fig6.5} Like Panels (d) and (h) in Figures \ref{fig5} and \ref{fig6},
    this figure shows the stripped fraction in the toy model versus that in the simulation,
    except here to be ``stripped'' means to exit the radius initially containing $99\%$ of
    the gas in the disk or halo, rather than to be energetically unbound. From left to right,
    the panels represent the timescales for $25\%$ and $90\%$ of the gas to be stripped from the ionized halo and
    $25\%$ and $75\%$ of the gas to be stripped from the neutral disk.}
\end{figure*}

We now check that the trends seen in Figures \ref{fig5} and \ref{fig6} are 
not somehow artifacts of the definition used for ``stripping''. Observations of gas stripping
in galaxies are generally based on physical separation of gas from the body of the galaxy, rather
than gravitational binding, which is not directly observable. Thus,
in Figure \ref{fig6.5}, we present results for stripping based on the assumption
that a particle is ``stripped'' when it exits the initial $99\%$ radius of either the 
neutral disk or ionized halo (depending on which it originally belongs to). We show a comparison
only between the toy model and simulation as the analytic model does not allow us to predict
the locations of particles.

The left panels of Figure \ref{fig6.5} show the $25\%$ and $90\%$ stripping timescales for the
ionized gas using the above definition of stripping, and the right panels show the $25\%$ and $75\%$
stripping timescales for the neutral gas. The trends are highly similar to those seen when
comparing the toy model to the simulation in Figures \ref{fig5} and \ref{fig6}. There is good
agreement for the $25\%$ stripping timescale of the ionized gas, but the $90\%$ timescale is
significantly underestimated by the toy model. In contrast, the $25\%$ stripping timescale 
for the neutral gas is overestimated by the toy model, and there is significant scatter between
the toy model and simulation predictions of the $75\%$ stripping timescale for the neutral gas,
although the median deviation between the two is small (59 Myr).

In this subsection, we have examined the differences in the stripping timescales
predicted by the analytic ram pressure stripping models, our toy model, and the EAGLE simulation.
For the $25\%$ stripping timescale of the ionized gas, which is often short (a few hundred Myr), 
we find that the toy model agrees with the EAGLE simulation, but the analytic model does not. The latter
is due to the fact that the definition of stripping in the analytic model differs from that in the simulation, 
and also because the analytic model somewhat underestimates the stripping timescale relative to the toy
model even using the same definition of stripping.

However, for the $90\%$ stripping timescale of the ionized gas, as well as both the $25\%$ and $75\%$
stripping timescales of the neutral gas, the above two factors become negligible, and the differences
between the analytic model and simulation are paralleled in the differences between the 
toy model and simulation. Both the analytic and toy models predict $90\%$ stripping timescales that are too
short for the ionized gas, $25\%$ stripping timescales that are too long for the neutral gas,
and $75\%$ stripping timescales for the neutral gas that have a significant scatter between their values 
and those of the simulation. We will explore the reason for these seemingly disparate
trends in the next subsection.

\section{Toy Model Deviations from the Simulation}\label{results_toy}

\begin{figure*}[t]
    \centering
    \includegraphics[width=\linewidth]{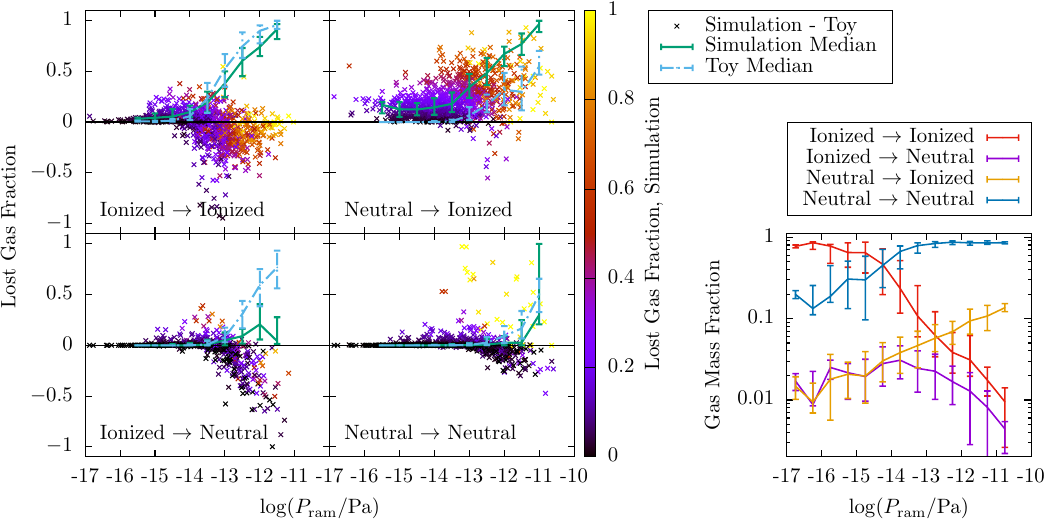}
    \caption{\label{fig7} \textit{Left four panels:} The fraction of gas mass unbound from
    each galaxy in the toy model and simulation between two simulation outputs (typically $\approx 120$ Myr), 
    as well as the difference between the two models. The top left panel shows these values for
    gas that is ionized at both the presented and subsequent simulation output (denoted ``II''), 
    the top right panel presents 
    gas that is initially neutral but becomes ionized (``NI''), the bottom left panel shows ionized gas that
    becomes neutral (``IN''), and the bottom right panel shows gas that is neutral and remains so (``NN''). 
    In each panel, the solid green 
    line denotes the median and quartiles of the distribution of unbound gas fraction
    in the simulation as a function of instantaneous ram pressure. The light blue dashed line is the same but
    for the prediction of the toy model. The points in each panel represent the unbound fraction from the
    simulation minus that from the toy model for each output for each galaxy. A simulation output is only
    included in this figure if a galaxy has at least 20 particles of a given type at that time. 
    The color of each point represents the unbound fraction from the simulation for that output and galaxy.
    \textit{Right panel:} The mass fraction of each category of gas 
    as a function of ram pressure: the red line is II gas, purple is IN, orange is NI, and blue is NN. 
    Lines denote median values and error bars denote quartiles.}
\end{figure*}

In this section, we examine the hydrodynamical and subgrid effects that
are present in the EAGLE simulation but not the toy model, and cause the
latter to deviate from the former. The important factors that we identify 
are stellar feedback, the cooling of ionized gas into denser neutral gas, 
and the fact that the gas appears to be confined by an additional
hydrodynamical force that increases with ram pressure.
As shown in \S\ref{results}, the differences in stripped gas fraction between the 
toy model and simulation also constitute the majority of the difference between the results
of the analytic models and those of the simulation.

Unlike in \S \ref{results_analytic}, in this section we show the results of the toy model
run over the time between each two subsequent simulation outputs (typically $\approx 120$ Myr). This
is summarized in Table \ref{table:results2} as the ``single-step'' toy model.
We do this because the toy model is simplified and will thus diverge from the simulation over time,
and so examining the difference between the toy model and simulation
over shorter timescales makes it easier to identify the physical processes
in the simulation that cause it to differ. Our ``single-step'' analysis 
results in a total of 1383 outputs for the
80 galaxies in our sample, each of which is treated as an independent
data point.

While we once again divide the
gas into neutral and ionized components, we now do this at each output rather
than only when we begin tracking the galaxy.
Furthermore, the ability of gas to transition between phases constitutes one of the important differences
between the simulation and toy model. We thus divide gas into four categories
in this section rather than two: gas that is neutral in the subsequent simulation output,
which we denote ``NN'' in the text for brevity; gas that is ionized 
in both simulation outputs (``II''); gas that is initially
neutral and becomes ionized in the next simulation output (``NI''); and gas that
transitions from ionized to neutral (``IN'').

In the rightmost panel of Figure \ref{fig7}, we show the gas mass fraction in each of 
these four categories per individual output as a function of ram pressure. 
The lines with error bars denote the median and interquartile range. 

At negligible ram pressures ($P_{\mathrm{ram}} \lesssim 10^{-15}$ Pa), when the
galaxies first begin their journey into the cluster,
the majority ($\approx 75\%$) of the gas bound to each galaxy is ionized and remains so (II), while
the next largest fraction of gas ($\approx 25\%$) is neutral and remains so (NN).
The fraction of II gas drops to only a few percent at high ram pressures 
($P_{\mathrm{ram}} \gtrsim 10^{-12}$ Pa), while
NN gas becomes the dominant type of gas bound to the galaxy ($\approx 90\%$ mass fraction). 

Gas that is ionized and becomes neutral (IN) is 
generally on the order of $1\%$ of the gas mass at all ram pressures.
However, over the full evolution of the galaxy in cluster, 
a substantial fraction of the initially ionized gas can be converted to neutral
gas. We find that a median of $33\%$ of the initially ionized gas becomes neutral at some
point (interquartile range $17\%$ to $50\%$), 
although some of this gas may become ionized again later. This plays a role in why 
the toy model overpredicts the stripping of the initially ionized gas, as seen if Figure \ref{fig5}. 
Some of this ionized gas cools to become denser neutral gas, which is more difficult to strip.

By contrast, gas that is initially neutral but becomes ionized (NI)
increases in the fraction of total mass it comprises from $\approx 1\%$ at low ram pressures to
$\approx 10\%$ at the highest ram pressures. 
Like for the IN gas, there is significant conversion over time 
from neutral to ionized gas: the median fraction of 
initially neutral gas converted to ionized gas at some point is $51\%$ (interquartile range
$41\%$ to $58\%$). NI gas is associated with stellar feedback, 
which we will discuss further is \S\ref{feedback}.

In the left four panels of Figure \ref{fig7}, we show, for each of these four categories of gas,
the single-step unbound gas fraction as a function of ram pressure. 
The green solid lines represent the median and quartiles of the unbound
gas fraction in the simulation. The same is shown for the toy model as light blue dashed lines.
The points show the difference between the unbound fraction in the simulation and toy model 
for each galaxy at each output, with the color representing the stripped fraction in
the simulation.

The top left panel of the left side of Figure \ref{fig7} plots the unbound fraction of 
II gas. We see that the toy model gives a similar increasing trend with ram pressure as the simulation, 
but predicts slightly too much 
stripping (by $\lesssim 10\%$) at ram pressures $\gtrsim 10^{-13}$ Pa. The IN gas
is shown in the lower left panel, and in contrast to the II gas, the toy model significantly overpredicts
the unbound gas fraction at $\gtrsim 10^{-13}$ Pa. As noted above, this is partly due to the fact
that this ionized gas cools and becomes denser --- a hydrodynamical
effect that is not included in the simple toy model. Both of these effects are in agreement with 
the results found for the ionized gas in the 
long-timescale integration shown in Figure \ref{fig5}, in which the toy model 
at low ram pressures matches the simulation, but tends to underpredict stripping timescales
at high ram pressures.

The top right panel in the collection of four panels seen on the left of Figure \ref{fig7} 
shows the NI gas. Here, the simulation stripped
fraction is larger than that from the toy model at all ram pressures, even negligible ones. 
As we will discuss further in 
\S \ref{feedback}, this is consistent with the effect of stellar feedback, which is implemented as 
stochastic and thermal in EAGLE.

Finally, the lower right panel of the left portion of Figure \ref{fig7} shows the NN gas. 
As noted above, this category of gas is dominant at 
high ram pressures, where most of the ionized gas has already been stripped. The NN gas
gas generally has a low stripping rate even at the highest ram pressures, although there are galaxies
where nearly all the gas is stripped in the simulation (appearing as yellow points). 
At high ram pressures ($P_{\mathrm{ram}} \gtrsim 10^{-12}$ Pa), the toy model overpredicts
the median stripped fraction of this category of gas, but the interquartile range of the stripped fraction
is significantly larger in the simulation. In the next two subsections, we will argue that this is due
to the combined effect of feedback from stars, and the action of a confining force on the galactic gas.

\subsection{Effect of Stellar Feedback} \label{feedback}

One factor that can contribute to excess gas loss in the simulation, particularly in the neutral gas disk,
is feedback. We find that in our galaxy sample, there is little supermassive black hole growth
compared to star formation, and the thermal energy added to the galactic gas correlates
well with the number of star particles formed but not with the black hole accretion rate.
Stellar feedback appears to be the main feedback source for galaxies in our sample,
and we thus focus here on feedback from newly formed star particles.

In EAGLE, feedback is implemented as purely thermal and stochastic,
such that some gas particles around a newly-formed star particle have
their temperature raised by $10^{7.5}$ K. The pressure gradient generated by
this influx of energy then drives a wind in the gas \citep{pmitchell}. 
Thus, the regions around newly formed star particles should exhibit a significant
excess of kinematically unbound particles. The thermal energy should also be
transferred among nearby gas particles and cause some neutral particles
to become ionized. Feedback should also act even at low ram pressures,
as star formation is present in the absence of ram pressure. This could
therefore account for the trend seen in the NI gas in Figure \ref{fig7},
whose rate of kinematic unbinding (which does not take into account thermal energy)
is underestimated by the toy model even at negligible ram pressures.

\begin{figure*}
    \centering
    \subfloat{\includegraphics[width=0.48\textwidth]{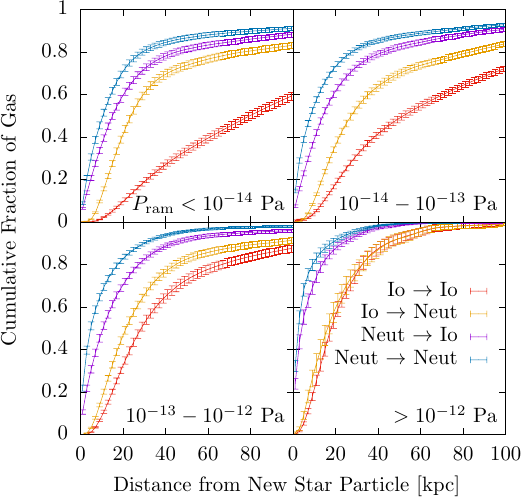}}
    \hfill
    \subfloat{\includegraphics[width=0.48\textwidth]{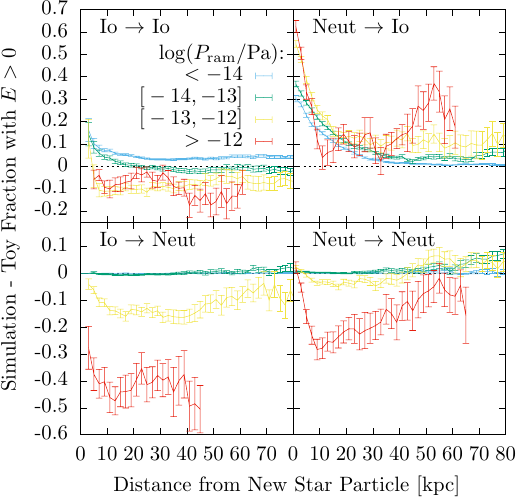}}
    \caption{\label{fig8} \textit{Left four panels:} The mean cumulative fraction of gas
    in EAGLE galaxies as a function of distance from newly-formed star particles.
    Different colored lines represent different categories of gas, with the colors and 
    categories being the same as for the right panel of Figure \ref{fig7}. The different
    panels represent bins of increasing ram pressure. Error bars are the error on the mean. 
    \textit{Right four panels:} The difference between the fraction of gas unbound 
    in the simulation and that unbound in the toy model, as a function of distance 
    from newly-formed star particles. The four panels represent the four different
    categories of gas. The colors represent increasing bins of ram pressure: light blue
    is $P_{\mathrm{ram}} < 10^{-14}$ Pa, green is $10^{-14}$ Pa $< P_{\mathrm{ram}} < 10^{-13}$ Pa,
    yellow is $10^{-13}$ Pa $< P_{\mathrm{ram}} < 10^{-12}$ Pa, and red is $P_{\mathrm{ram}} > 10^{-12}$ Pa.
    Error bars are the error on the mean. Bins in which fewer than 10 galaxies
    have any gas particles are not shown.}
\end{figure*}

We further examine the effects of stellar feedback in Figure \ref{fig8}.
In the left four panels, we compare the spatial distribution of the four
gas categories seen in Figure \ref{fig7}
as a function of average distance from newly-formed star particles in their galaxies.
This was computed by stacking all the new star particles expected to undergo feedback between simulation outputs,
and averaging the cumulative fraction of total gas around them as a function of distance. (The positions
used are the initial positions of all particles rather than evolved ones at the 
moment of feedback; thus the correlation may be decreased relative to its true strength.)
The stacked means are then averaged over all outputs/galaxies. Error bars represent the errors
on the mean.

The four panels on the left of Figure \ref{fig8} represent
bins of different ram pressure, while the colored lines represent the four gas categories (II, NI, IN, and NN).
We see that at all ram pressures, the initially neutral gas --- 
both NN and NI gas --- is closer to regions
that are forming stars than the initially ionized gas. 
This is expected, as stars form in dense gas in EAGLE.
The IN gas is closer to the star-forming regions than the II gas, 
particularly at low ram pressures. This is likely because the
ionized gas that becomes neutral is that closest to the neutral gas disk,
and cools onto it. Overall, we would expect a larger fraction of the total neutral gas to be affected
by stellar feedback than the ionized gas, simply due to 
the greater proximity of the neutral gas to newly-formed star particles.

We also note that both the neutral and ionized gas distributions become closer to newly-formed
star particles on average with increasing ram pressure. This is partly because diffuse gas far from
the neutral disk is already stripped at high ram pressures, but also partly because 
the force of ram pressure tends to compress gas --- which also triggers increased star formation \citep{troncoso}.

In the four right panels of Figure \ref{fig8}, we plot
the mean difference between the unbound gas fraction in the simulation and that predicted by the toy model 
 as a function of distance from newly-formed star particles. 
The average is computed in the same manner as for the cumulative fraction of gas in the left
panels, described above. Here the different panels show the four different categories of gas,
while the different line colors indicate bins of different ram pressure. Since feedback in EAGLE 
is implemented as a temperature increase in gas particles close to newly-formed star particles, we would 
expect feedback-related differences between the simulation and toy model to be correlated
with the locations of the newly-formed star particles.

The top right panel shows NI gas, which we would expect
to result from the thermal energy that is injected by feedback in EAGLE. We emphasize that the unbound
gas fractions presented in this figure are based solely on the sum of the kinetic and potential energy 
of the gas particles, and do not include the thermal energy. We see that NI gas particles
close to newly-formed star particles are significantly 
more likely to have a velocity higher than the escape speed in the simulation than in the toy model.
The correlation with the location of newly-formed star particles is evident in all ram pressure bins.
Furthermore, at higher ram pressures, the difference in unbound fraction
between the simulation and toy model is larger. 
This implies that ram pressure assists feedback in expelling gas from galaxies.

In the bottom right panel of the four right panels of Figure \ref{fig8}, 
we plot the same quantity described above
but for the NN gas. Here there is little difference between the simulation and toy model
for $P_{\mathrm{ram}} \lesssim 10^{-13}$ Pa. As seen in Figure \ref{fig7}, 
at these low ram pressures, very little NN gas becomes unbound in either model.
At higher ram pressures, however, we see two effects. The difference between
the simulation and toy model unbound fraction is enhanced relative to the baseline value 
closer to newly-formed star particles, but the overall stripped fraction in the simulation is lower than that in 
toy model. We will discuss the reason for the latter in \S\ref{confinement}. %SOMETHING
Nevertheless, the fact that Figure \ref{fig8} is an average over all
galaxies suggests that it is consistent with Figure \ref{fig7} for the NN gas: some 
galaxies at high ram pressure have NN gas that is significantly unbound by feedback, 
whereas in others the overall underprediction of stripping by the toy model relative to the simulation 
is the dominant effect.

In the lower left panel of the right set of panels in Figure \ref{fig8}, we see the same types of curves
as described above but for IN gas. Like for the NN gas,
at $P_{\mathrm{ram}} \lesssim 10^{-13}$ Pa there is no significant difference between the 
simulation and toy model. This is also visible in Figure \ref{fig7}, where both models generally
predict no stripping at these ram pressures. At higher ram pressures, we see similar trends as for
the NN gas, but there is an even larger overall negative
difference in the stripped gas fraction between the simulation and the toy model. 
This is because the ionized gas becomes more dense when it
becomes neutral, which makes it more resistant to stripping, an effect not 
accounted for in the toy model. We will discuss this in more detail in \S\ref{confinement}.

Finally, in the top left panel of the right half of Figure \ref{fig8}, we show the results for the II gas.
Here the unbound fraction is again higher than the 
baseline close to newly-formed stars (except in the highest ram pressure bin, for which there
is not enough data at small distances). Similar to the IN and NN gas, the overall level
of stripping in the simulation relative to the toy model decreases with increasing ram pressure. In the 
lowest ram pressure bin, there is a constant low fraction of lost ionized gas that may be
due to stochastic unbinding of these very weakly bound gas particles.

Although the II gas and IN gas are locally affected by feedback at 
small distances from newly-formed star particles, 
they will be less affected overall than the NN and NI gas. This is because
the majority of all II and IN gas is located
at larger distances from stellar feedback regions than NN and NI gas, 
as can be seen in the left half of Figure \ref{fig8}.

The differences between the simulation and toy model for the different categories of gas presented 
in Figure \ref{fig8} explain a number of the trends seen in Figure \ref{fig7}.
The NI gas is the category of gas that is most strongly affected by stellar feedback,
and thus is stripped more in the simulation than in the toy model at all ram pressure values. 
For the other gas categories, the simulation sometimes predicts more stripping
than the toy model, but more often it predicts less stripping, especially at high ram pressures. This can be seen
in Figure \ref{fig8} as a negative difference between the simulation and toy model stripped gas fractions
far from regions that are undergoing stellar feedback. In the following section,
we will discuss the reasons that the simulation predicts less stripping than the toy model at high ram pressures.

\subsection{Influence of Gas Mass Surface Density} \label{confinement}

\begin{figure}[t]
    \centering
    \includegraphics[width=\linewidth]{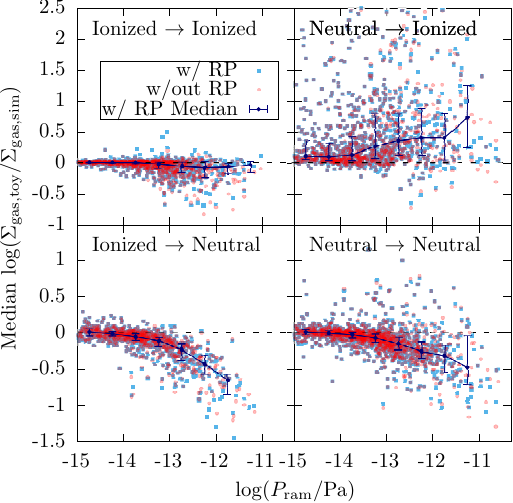}
    \caption{\label{fig9} The median ratio between the toy model and simulation predictions for the gas mass
    surface density $\Sigma_{\mathrm{gas}}$ perpendicular to the ram pressure, 
    as a function of ram pressure. Each point represents a single galaxy between two subsequent simulation
    outputs. The light blue points show the ratio 
    predicted by the toy model with ram pressure included, 
    while the red points are for the toy model integration assuming zero ram pressure. 
    The dark blue lines represent the median and quartile values of the distributions
    of the light blue points.
    In each panel, the only outputs shown are those in which the 
    galaxy has at least 20 particles of the given gas category. The values
    of $\Sigma_{\mathrm{gas}}$ are computed using all of the gas particles bound to the subhalo
    at each starting output, regardless of whether they are unbound by the subsequent output. 
    The four different panels present the same four gas categories shown in Figures \ref{fig7}
    and \ref{fig8}.}
\end{figure}

We have seen in Figure \ref{fig8} that, away from galactic regions strongly affected by stellar feedback,
the toy model increasingly overpredicts the fraction of unbound gas with increasing ram pressure. 
A significant cause of this is underprediction of the gas mass surface density by the toy model. 
This is presented in Figure \ref{fig9}, which shows the ratio of 
toy model to simulated predicted gas mass surface density, $\Sigma_{\mathrm{gas}}$,
perpendicular to the ram pressure direction.
The toy model prediction uses the positions and velocities of the galaxy gas particles
in each simulation output,
as well as the ram pressure and subhalo potential,
to predict their positions in the next simulation output.
This predicted gas distribution is used to calculate the toy model $\Sigma_{\mathrm{gas}}$
at the location of each gas particle. 
The simulation $\Sigma_{\mathrm{gas}}$ is computed from the true final positions of the gas particles in
the simulation. The ratio between the toy model and simulation $\Sigma_{\mathrm{gas}}$ for each output/galaxy
is shown as the light blue points in Figure \ref{fig9}, 
while the median and quartiles of their distribution in bins of 
ram pressure is represented by the dark blue lines.
We note that $\Sigma_{\mathrm{gas}}$ is computed using all of the gas present in the galaxy
in the initial simulation output, regardless of gas phase: each gas component
contains the density contribution of the other components that are in its line of sight.

Different behavior can be seen for different categories of gas in Figure \ref{fig9}.
For the NI gas, seen in the top right panel of Figure \ref{fig9}, 
the toy model nearly always predicts higher gas mass surface densities than the simulation. 
This is unsurprising, as the NI gas is associated with stellar feedback events,
which are not accounted for in the toy model and induce a wind in the gas, causing the gas mass surface
density to become lower.

For the II gas in Figure \ref{fig9}, we see that the median ratio
between the toy model gas surface density and that in the simulation tends to be close to one, 
although at high ram pressures, some galaxies have significantly lower densities in the toy model
than in the simulation. The latter effect is even more pronounced for the NN and
IN gas, for which the toy model strongly underpredicts the gas mass surface density at high
ram pressures.

We emphasize that none of the trends seen in Figure \ref{fig9} 
are the result of differing amounts of gas being
stripped in the toy model versus the simulation. This is because the values of $\Sigma_{\mathrm{gas}}$ used
to create Figure \ref{fig9} are computed using all of the gas particles that are
initially bound to the subhalo, regardless of whether
they have become unbound in the subsequent simulation output. Thus, the difference in $\Sigma_{\mathrm{gas}}$
is due to the different final positions of the gas particles predicted by the toy model
and simulation, rather than due to particles being excluded due to becoming unbound. For the NN
and IN gas especially, the toy model predicts a gas distribution that is more diffuse in the direction perpendicular
to the ram pressure than the simulation.

\begin{figure*}[t]
    \centering
    \includegraphics[width=\linewidth]{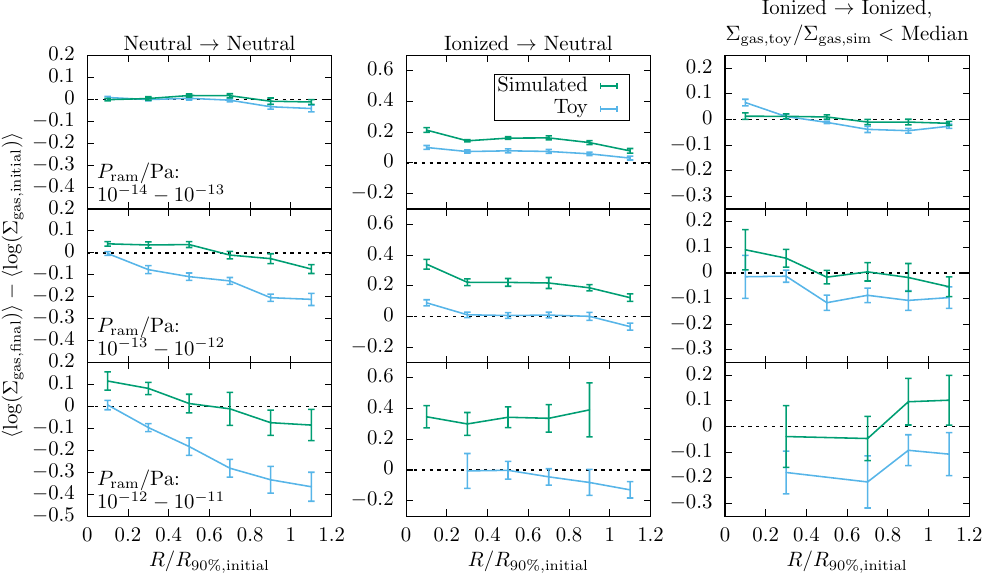}
    \caption{\label{fig10} The evolution of $\Sigma_{\mathrm{gas}}$, the gas mass surface density perpendicular
    to the ram pressure, as a function of 
    the radius perpendicular to the ram pressure $R$, scaled by the $90\%$ value of $R$ at the 
    initial simulation output (computed separately for the different categories of gas). 
    The evolution of $\Sigma_{\mathrm{gas}}$ is expressed as the difference between 
    the final and initial mean $\log(\Sigma_{\mathrm{gas}})$ of the gas particles in each radial bin for each galaxy. 
    This is averaged over all the galaxies in a bin of ram pressure.
    The green lines represent the evolution of $\Sigma_{\mathrm{gas}}$ from the EAGLE simulation, whereas the light blue lines
    represent the evolution predicted by the toy model without ram pressure (see text). 
    The rows indicate bins of increasing ram pressure, while the 
    columns represent three categories of gas: NN, IN, and II. The II gas is averaged over only
    the galaxies that fall below the median relation of $\Sigma_{\mathrm{gas, toy}}/\Sigma_{\mathrm{gas, sim}}$
    as a function of ram pressure. Galaxies/outputs included in 
    the average are those with at least 20 gas particles of the indicated type. 
    Error bars represent errors on the mean.}
\end{figure*}

The results seen in Figure \ref{fig9} are similar to those seen in Figure \ref{fig4}, 
which presented the evolution of $\Sigma_{\mathrm{gas}}$ for only the bound particles
remaining at each time during the long-timescale integration.
In Figure \ref{fig4}, $\Sigma_{\mathrm{gas}}$ of both the initially ionized and initially neutral
gas became higher in the simulation than in the toy model. In the simulation, some of the initially
ionized gas will become neutral, and therefore denser and
less likely to be unbound than gas that remains ionized. From this alone, we would expect
the initially ionized gas that remains in the galaxy to be denser than is predicted
from the toy model, because some of it has since become neutral. 
However, in addition to this, once the gas has become neutral, 
NN gas is more dense in the simulation than predicted by the toy model,
as seen in Figure \ref{fig9}. 
Conversely, NI gas becomes less dense, but is more likely to be stripped from
the galaxy than initially neutral gas that remains neutral. Thus it is expected
that both the initially ionized and initially neutral gas that remain bound
to the galaxy are denser in the simulation than in the toy model.

Because ram pressure moves gas to the outer, less bound regions
of the subhalo, it can cause the distribution of gas to expand in the perpendicular direction. 
Thus, adopting an overly large value for the ram pressure in the toy model could theoretically cause
the gas surface density in the toy model to be lower than in the simulation.
However, we demonstrate that this \textit{not} the cause of the discrepancy
via the red points in Figure \ref{fig9}. These show the ratio of the toy model 
to simulated $\Sigma_{\mathrm{gas}}$ for a toy model in which the ram pressure is set to zero:
the gas particles are simply evolved from their initial positions and velocities under the
gravitational potential of the subhalo. We see that the values of $\Sigma_{\mathrm{gas}}$ predicted by
this model are still smaller than the values from the simulation.

Since the gas in simulated galaxies is more compact than predicted by
the toy model, it will be more resistant to ram pressure than expected, 
leading to lower stripped gas fractions.
This is a significant cause of the lower stripping
rate in the simulation compared to the toy model noted in Figures \ref{fig7} and \ref{fig8} 
at high ram pressures for II, NN, and IN gas.

We would like to further examine the different predictions made 
for the gas mass surface density $\Sigma_{\mathrm{gas}}$ in the toy model and simulation. 
We thus present the average evolution of $\Sigma_{\mathrm{gas}}$ 
between two subsequent simulation outputs for the NN, IN, and II gas in Figure \ref{fig10}.
The evolution is shown as the difference between the final and initial mean $\log(\Sigma_{\mathrm{gas}})$, 
for both the toy model (without ram pressure) and the simulation, 
as a function of radius perpendicular to the ram pressure (scaled by the 
initial $90\%$ radius). The initial and final mean values of $\log(\Sigma_{\mathrm{gas}})$ 
are computed as an average over all the gas particles in each radial bin, using 
all gas particles that were bound to the galaxy at the initial simulation output.
The values of $\Sigma_{\mathrm{gas}}$ are thus averaged
per gas particle, rather than per unit area, meaning that an increase in $\Sigma_{\mathrm{gas}}$ at all radii
indicates an increase in the clustering of gas particles at all radii.

We associate IN gas with an increase in density due to cooling, and this is indeed what
is seen in the center column of Figure \ref{fig10}. The toy model $\Sigma_{\mathrm{gas}}$ does not
evolve significantly between two simulation outputs, but $\Sigma_{\mathrm{gas}}$ in the simulation increases
at all radii. The disparity between the simulation and toy model increases with ram pressure.

By contrast, in the left column of Figure \ref{fig10} showing the NN gas,  
$\Sigma_{\mathrm{gas}}$ from the simulation does not evolve significantly, while $\Sigma_{\mathrm{gas}}$ in
the toy model tends to decrease between two simulation outputs. The discrepancy between
the simulation and toy model increases with radius and with ram pressure. 
This trend suggests a confining force acting on the gas
whose strength is correlated with the ram pressure.

While most outputs/galaxies show little difference between the toy model
and simulation for the predicted $\Sigma_{\mathrm{gas}}$ of the II gas, 
we would like to examine the evolution of $\Sigma_{\mathrm{gas}}$ for
those outputs where there is a discrepancy between the two. 
Thus, the right column of Figure \ref{fig10} shows the II gas, but only for those outputs/galaxies
that fall below the median trend of $\Sigma_{\mathrm{gas, toy}}/\Sigma_{\mathrm{gas, sim}}$ with ram pressure. 
Here the II gas demonstrates a trend more similar to the NN gas than the IN gas: 
$\Sigma_{\mathrm{gas}}$ does not evolve significantly in the simulation, but tends to decrease in the toy model.

Figure \ref{fig10} suggests a confining force on the galaxy gas that acts in addition
to the force of the gravitational potential of the subhalo, which is accounted for 
in the toy model. We find that the self-gravity of the galaxy gas, which is not accounted
for in the simple toy model, is not sufficient to explain the additional
force. Nor is the force caused by unaccounted-for changes in the subhalo potential, 
as we have run our toy model integration (without
ram pressure) on the collisionless stellar particles, finding that their
predicted positions are in good agreement with those from EAGLE.
This implies that the confining force seen in Figure \ref{fig10} is 
hydrodynamical in nature. While the thermal confinement pressure
of the surrounding ICM could potentially produce the observed confinement effect, we do not find
a correlation between $\Sigma_{\mathrm{gas, toy}}/\Sigma_{\mathrm{gas, sim}}$
and the ICM confinement pressure at fixed ram pressure, although there is
a correlation between $\Sigma_{\mathrm{gas, toy}}/\Sigma_{\mathrm{gas, sim}}$
and ram pressure at fixed confinement pressure. This suggests that
the observed confining force may be somehow related to the ram pressure force.

Additionally, we have run our toy model without ram pressure on a sample of cluster
galaxies from the Illustris TNG 100 Mpc simulation, chosen in a similar manner as our
EAGLE galaxy sample. We find the same result regarding the galactic gas being
more compact than expected at high ram pressure that we see in EAGLE. Therefore,
this effect does not appear to result from the specific hydrodynamic scheme used in EAGLE.

\begin{figure}[t!]
    \centering
    \includegraphics[width=\linewidth]{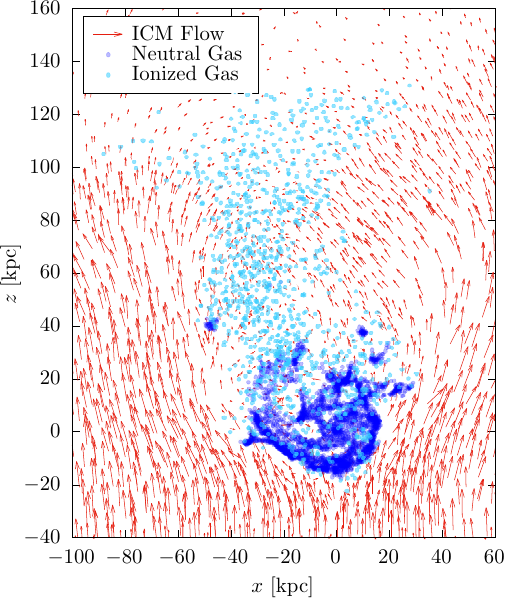}
    \caption{\label{fig11} The flow of ICM gas particles around a galaxy in our sample. The ICM particles
    presented are those in a slice between the tenth and ninetieth percentile y-axis coordinates of the galaxy gas
    distribution, where the y-axis is that perpendicular to the x-z plane displayed in the figure. The z direction
    is the overall ram pressure direction, computed as described in \S \ref{analytic}. Each ICM particle
    in the slice is plotted as a red arrow proportional to the particle's velocity in the x-z plane. The 
    galaxy's neutral gas is plotted as dark blue points and the ionized gas as cyan points. It is apparent that
    the ICM flow is deflected around the galaxy gas.
    }
\end{figure}

As the difference in $\Sigma_{\mathrm{gas}}$ between the toy model and simulation 
correlates strongly with ram pressure, we suggest that its cause may be deflection
of the ICM wind by the galactic gas. The standard \citet{gunngott} formula for the ram pressure,
$P_{\mathrm{ram}} = \rho_{\mathrm{ICM}} v_{\mathrm{rel}}^{2}$, 
assumes that all of the momentum of the ICM wind is transferred to the galactic gas.
However, ICM particles tend to be deflected around the galaxy in
the simulation, such that they do not deposit all of their momentum into
the galaxy gas. Such a deflection will tend to produce an inward force on the galactic gas 
that increases $\Sigma_{\mathrm{gas}}$, and also to decrease the force exerted in the assumed
ram pressure direction, which will further decrease the stripping
rate relative to the toy model prediction. We would expect the magnitude
of the confinement from this effect to increase with ram pressure, although it
will also depend on the geometry of the galaxy gas. We demonstrate the deflection 
of ICM particles around an example galaxy in Figure \ref{fig11}.

\subsection{Discussion of Toy Model Deviations}

We have seen in this section that there are multiple hydrodynamical effects that
alter the stripping rate of gas from ram pressure stripped galaxies.
One is that gas is able to transition between neutral and ionized
and this affects a significant fraction of the initially bound gas.
Initially ionized gas can cool and become denser and more difficult to strip, as is 
demonstrated in the ``Ionized $\rightarrow$ Neutral'' panels in Figure \ref{fig9}
and \ref{fig10}. Similarly, neutral gas can become ionized and decrease in density,
seen in the ``Neutral $\rightarrow$ Ionized'' panel of Figure \ref{fig9}.
The fraction of total gas undergoing each of these two 
processes at each simulation output can be seen as a function of ram pressure in the 
rightmost panel of Figure \ref{fig7}.

The transformation of neutral gas to ionized is associated with stellar 
feedback, which significantly alters the stripping rate
by inducing a pressure-driven wind in the gas. In the right four panels of Figure \ref{fig8}, this 
can be seen as a larger local fraction of stripped gas in the simulation
than the toy model close to newly-formed star particles. This is most readily
seen in the top right panel for the neutral gas that becomes ionized, which we would expect to be associated with
the thermal feedback implementation in EAGLE. However, it is also visible
as an increase over the baseline difference between the simulation and toy model stripping
rate for all four categories of gas presented. The stripping enhancement caused by the feedback
becomes more pronounced with increasing ram pressure, implying that the ram pressure and feedback
work in tandem to unbind gas from the galaxy.

That ram pressure and stellar feedback work together to unbind more gas
than either would individually is in agreement with previous results obtained by \citet{feedback_gimic}
using the GIMIC simulation, even though the implementation 
of feedback in their simulation was kinematic rather than thermal.
That galaxies continue to form stars while they are being ram pressure stripped
is also consistent with results from the Illustris TNG simulations \citep{goller_illustris}.

We note, however, that our analysis likely overestimates
the impact of stellar feedback on ram pressure stripped galaxies, as it has been found that 
stellar feedback in EAGLE is too strong due to the limited resolution of the simulation. 
Because the mass of a star particle in EAGLE 
is much larger than a single star, there is too much feedback energy injected
into a small region of the galaxy, leading to unphysically large holes
in the neutral gas \citep{bahehi}. \citet{hydr} attribute to this the fact
that galaxies in the Hydrangea simulations, a set of resimulations of clusters
with the EAGLE physics implementation, show excessive quenching at stellar masses below
$10^{9.5} \msun$. In the right panels of Figure \ref{fig8}, we also see enhanced stripping
around the locations of newly-formed star particles out to tens of kpc.
Nevertheless, it is still likely that
stellar feedback makes some contribution to the effect of ram pressure stripping
in the real universe, especially given that compression of gas by
ram pressure has been reported to enhance star formation \citep{gaspsf} --- an effect also seen
 in the leading edges of ram pressure stripped galaxies in EAGLE \citep{troncoso}.

Enhanced AGN activity has also been observed in ram pressure stripped 
galaxies \citep{peluso}, but we do not find this in our sample. This may be because
EAGLE is unable to resolve the scales relevant to supermassive black hole feeding
and employs simple prescriptions for their accretion. 
Other simulations suggest that ram pressure can enhance accretion onto the 
central supermassive black hole \citep{romulusAGN, akerman}.

While we find evidence that stellar feedback significantly enhances the stripping
rate of gas from galaxies in EAGLE, we find an overall suppression of the stripping
rate in the simulation relative to the toy model that increases with ram pressure. 
As noted above, for gas that is initially ionized but becomes neutral, this can be explained by the
fact that the gas is cooling and thus becomes more compact and difficult to strip. This can be seen 
in the center column of Figure \ref{fig10}, which shows that this type of gas increases
in surface mass density between each two simulation outputs at all distances from the galaxy center,
an effect not present in the simplified toy model.
We note that the increase in gas surface density appears to be larger at higher ram pressures,
as can be seen in the three bins of increasing ram pressure shown in Figure \ref{fig10}.

However, the toy model also predicts gas mass surface densities that are too low
for neutral and ionized gas that do not transition to the other state by the subsequent
simulation output. Figure \ref{fig10} indicates that
these types of gas become excessively diffuse in the toy model.
This suggests that some confining force that
is not present in the simplified toy model acts on the gas. We suggest that
this effect may be caused by the deflection of the ICM around the galaxy,
creating an inwards force and decreasing the total momentum transferred
to the galaxy gas in the ram pressure direction.
We demonstrate the deflection of the ICM gas around an example galaxy
in Figure \ref{fig11}.

\section{Conclusions}\label{conclusions}

In the previous sections, we compared the evolution of the stripped gas mass fraction
of the neutral gas disk and hot ionized halo of galaxies,
as predicted by simple analytic models, a more complicated toy model that integrates
the motion of gas under ram pressure, and a full 
hydrodynamical simulation. The analytic models we tested were the formulation 
from \citet{gunngott} for the neutral gas disk and that from 
\citet{mccarthy} for the hot ionized gas halo. The details
of these analytic models, and of our toy model,
are summarized in \S\ref{models} and Table \ref{table:results2}. 

The galaxy sample consisted of 80 simulated galaxies
across eight clusters at $z = 0$ from the EAGLE simulation, with the evolution of
the galaxies being tracked from their entry into the cluster friends-of-friends group. 
The properties of the galaxies are summarized in Figures \ref{fig1} 
and \ref{fig2} as well as Table \ref{table:results1}. 

\begin{enumerate}[leftmargin=*]
\item We found that the initial conditions assumed by the \citet{mccarthy} model for the
gas surface density and maximum gravitational restoring acceleration were
in approximate agreement with the true initial surface density from the
simulation and the maximum restoring acceleration predicted from kinematically
evolving the gas particles in the potential.

\item For the \citet{gunngott} model, we found that the 
maximum gravitational restoring acceleration was significantly
underpredicted if taking the gravitational force to come only from
the stellar disk, but was much better approximated if accounting for
the dark matter by using the same gravitational restoring acceleration
as for the \citet{mccarthy} model (Figure \ref{fig3}).

\item These analytic models for ram pressure assume that the maximum
gravitational restoring acceleration experienced by gas and the gas mass surface density
perpendicular to the direction of the ram pressure do not evolve over time. However,
this is not the case in a realistic ram pressure stripping situation: the ram pressure
changes direction and magnitude as the galaxy moves through the cluster,
and the orbits of gas as well as the gas surface density will evolve.
We plot this evolution as predicted by our toy model, which takes into account the above factors,
in Figure \ref{fig4}. This shows a substantial
evolution in the gas surface density within the toy model, with the
neutral gas becoming less dense over time and the ionized gas becoming
slightly more dense on average, but with substantial scatter among galaxies.

\item The evolution of the gas surface density in the toy model for the ionized gas
is reflected in the difference between the toy model and analytic model predictions
for the stripping timescales of the ionized gas (Figure \ref{fig5}, Panels b and f). It is
also present in the difference between the analytic model and EAGLE simulation in Panels (a) and (e), 
suggesting that it has some true effect.
 
\item The majority of the difference between the stripping timescales
predicted by the analytic models and the simulation are also reflected in the difference
between the toy model and the simulation, for both the ionized (Figure \ref{fig5})
and neutral (Figure \ref{fig6}) gas. This suggests that the disparity between
the analytic models and the simulation is primarily driven by hydrodynamical and subgrid
effects that are present only in the simulation.

\item We identify three main processes in EAGLE that cause
the stripping rate to deviate from analytic prescriptions in opposing directions. 
The gas stripping rate is enhanced by stellar feedback. It is decreased by 
cooling of the initially ionized gas into denser neutral gas,
as well as by a confining force on both the neutral and ionized gas.

\item The excess fraction of gas mass expelled per stellar feedback event 
(above the prediction for ram pressure stripping alone) increases
with increasing ram pressure (Figure \ref{fig8}), implying that 
ram pressure and feedback work synergistically to expel more gas than would
be expected merely from the sum of the two effects.

\item A median of $33\%$ of the gas that is initially in the ionized halo
of our galaxies becomes neutral at some point during their evolution 
(though it may become ionized again later). This enhances the density of the gas
in a way that cannot be predicted by simplified models (Figure \ref{fig10}, center column), 
making it more resistant to stripping.

\item  We find that there seems to be a confining force on the gas in our galaxy sample
that is stronger at higher ram pressure and leads to the gas
being less diffuse than expected from its kinematic initial conditions (Figures \ref{fig9} and \ref{fig10}). 

\end{enumerate}

Overall, while we find that the toy model improves upon the predictions of simple analytic models
in limited circumstances (in particular, for the $25\%$ stripping timescale of the ionized halo
shown in Figure \ref{fig5}), the results are otherwise similar to those of analytic models
when assuming realistic initial conditions for the latter. This suggests that only small 
improvements can be made to semi-analytic models 
by adopting more complex prescriptions for ram pressure stripping itself.

Instead, the majority of the disparity between 
the predictions of the EAGLE simulation and those of both the analytic and toy models 
results from hydrodynamical and subgrid effects. It therefore may be necessary to incorporate
prescriptions for such effects in semi-analytic models of ram pressure stripping. However,
more comparisons between hydrodynamical simulations and observations of ram pressure
stripped galaxies are also needed, as different simulations vary in resolution, hydrodynamical scheme,
and subgrid prescriptions, and therefore potentially also in their results.

\begin{acknowledgments}
This project has received funding from the European Research
Council (ERC) under the European Union's Horizon
2020 research and innovation programme (grant
agreement No. 833824).

This work used the DiRAC@Durham facility managed by the Institute for Computational Cosmology 
on behalf of the STFC DiRAC HPC Facility (www.dirac.ac.uk). The equipment was funded by BEIS 
capital funding via STFC capital grants ST/P002293/1, ST/R002371/1 and ST/S002502/1, Durham 
University and STFC operations grant ST/R000832/1. DiRAC is part of the National e-Infrastructure.

AI acknowledges the INAF founding program 'Ricerca Fondamentale 2022' (PI A. Ignesti). 
RS acknowledges financial support from FONDECYT Regular 2023 project No. 1230441.
\end{acknowledgments}

\appendix
\restartappendixnumbering

\section{Effect of Tidal Forces} \label{appendix}

\begin{figure*}
    \centering
    \includegraphics[width=\linewidth]{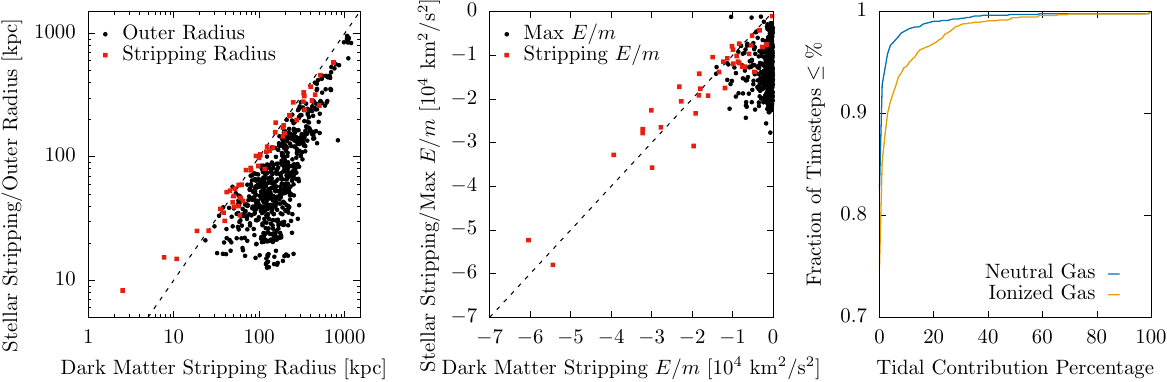}
    \caption{\label{figappendix} 
    \textit{Left:} The estimated tidal truncation radius of the dark matter, computed as described in the text,
    versus the estimated tidal truncation radius of the stars (red squares), or for galaxies in which the latter is non-existent, the maximum radius of the stellar distribution (black circular points). \textit{Center:}
    Similar to the left panel, but for the specific binding energy
    above which particles are tidally unbound, where the
    black points show the maximum specific binding energy
    for the stellar distribution. \textit{Right:} The
    fraction of simulation outputs having an
    estimated tidally stripped fraction of gas
    at or below the percentage given on the horizontal axis. The blue line 
    represents the neutral gas component and the orange line represents the ionized gas.}
\end{figure*}

In addition to ram pressure, galaxies in clusters
experience tidal interactions with both the central potential of the cluster
and other satellite galaxies. A statistical study of these processes in 
the EAGLE simulation was presented in \citet{marasco}.
Unlike ram pressure, tidal forces can strip stars and dark matter from galaxies in addition to gas.
To limit the number of tidally-affected galaxies in our sample,
we did not select galaxies that decreased in stellar mass during the
EAGLE snapshot in which they entered their host clusters.
However, the galaxies are nevertheless able to lose stellar mass during
their later evolution. It is also possible for them to lose stellar mass to tidal forces 
in their outer regions while gaining a larger amount in their inner regions
due to star formation. Furthermore, the ionized gas halo and sometimes
the neutral gas disk extend further out in radius than the stellar distribution,
and thus can be affected by tides even if the stars are unaffected.
For these reasons, we attempt here to estimate the maximum possible
effect of tides on our simulated galaxy sample (\S \ref{galsamp}).

We exploit the fact that tidal forces, whether from the central galaxy or other satellites, affect
stars and dark matter as well as gas. Continuous tidal stripping, such as that typically experienced 
by a galaxy traveling around the center of a cluster, produces
a tidal radius outside of which matter is stripped. This radius should be
equal for all types of matter within the galaxy. At each simulation output, we find the dark matter and stellar
radii outside of which $90\%$ of the particles of that type become unbound, 
as long at there are at least 30 particles outside the radius.
Defined this way, such a radius does not always exist, particularly for the star particles,
because the star particle distribution is less physically extended that the dark matter distribution
and thus requires stronger tides to be tidally stripped.
We consider the above quantity an estimate of the potential tidal truncation radius.

We would expect this method to fail to estimate the truncation radius 
of the galaxy gas in cases where the gas extends further than both the dark matter and the stars, and neither
the dark matter nor the stars have a well-defined truncation radius. If the spatial extent of the different
particle types is taken to be their $99\%$ radii, then it
is only for $4\%$ of the outputs in our sample that the truncation radius is not successfully estimated.

We present the estimated truncation radius in the leftmost panel of Figure \ref{figappendix}. 
In cases where the radius exists for both the dark matter and stars, 
we plot the two against each other as red squares. 
Here we see that the estimated truncation radius is generally similar for both
types of particles, as would be expected for the tidal radius. Black round points represent
galaxies for which the radius exists for the dark matter but not the star particles.
For these galaxies, we have
instead plotted the truncation radius for the dark matter versus the
outer edge of the stellar distribution, represented by the radius of the thirtieth most
distant star particle. The black points show that the edge of the stellar distribution in
these galaxies is typically smaller
than the truncation radius of the dark matter, as would be expected if the tidal
radius is outside the stellar distribution but within the dark matter distribution.

At very small stripping radii, we do see a discrepancy between the tidal
radii implied by the stars and those implied by the dark matter. 
This is because the mechanism of tidal stripping
is not always continuous stripping defined by a tidal radius, but is sometimes
due to tidal shocks that occur during rapid tidal interactions, such as those between
satellite galaxies in the cluster. These interactions unbind matter from the galaxy
by increasing its energy. The dark matter is less bound
than the stars and this mechanism therefore 
strips it to a smaller radius; this has been noted in EAGLE \citep{jingtidal}.

To estimate the amount of stripping if the mechanism is tidal shocks, we repeat
the calculation described above for the tidal radius, but instead compute the
binding energy per unit mass above which $90\%$ of stellar or dark matter particles are stripped. 
In the center panel of Figure \ref{figappendix}, we plot this value for
the dark matter versus the stars as red squares. The simulation outputs at very negative values 
of the plot correspond to the points in the left panel at small stripping radii, showing
that the energetic increase of the stars and dark matter is similar but
the dark matter is stripped to a smaller radius. As in the left panel,
in many galaxies the stars are not stripped, so we plot as black points
the limiting specific binding energy of all the stars, represented by the value for the
thirtieth least bound star particle. We see from this 
that the stars are typically more bound than the dark matter.

While ram pressure does not directly affect the stars and dark matter as
tidal forces do, it indirectly decreases how bound they are by 
removing gas mass from the subhalo \citep{smithtidal}.
Therefore, the stripping parameter estimates shown in the left and center panels of Figure
\ref{figappendix} likely overestimate the influence of tides on the stars and dark matter.

Nevertheless, we estimate the fraction of gas that is possibly stripped by tides 
by taking the maximum of four values: the fraction of gas outside the estimated
 tidal radius of the dark matter, the fraction of gas with specific binding energy
larger than the limiting specific binding energy of the stripped dark matter, 
and both of the above quantities but for the stars.
We present the estimated fraction of neutral and ionized gas stripped by tides 
in the right panel of Figure \ref{figappendix}.
This is shown as a cumulative histogram of the fraction of simulation outputs
(for all galaxies) that have an estimated tidally stripped mass fraction lower than
some percentage. The blue line represents the neutral gas and the
orange line the ionized gas. As might be expected, the ionized halo, 
which is more radially extended and less strongly bound, is more affected by tides than the
neutral gas disk. However, for both gas phases, the majority of simulation outputs have little expected
influence from tides.

The right panel of Figure \ref{figappendix} shows the \textit{instantaneous} fraction of gas mass
we predict to be stripped by tides at each simulation output, but we also compute 
the expected cumulative fraction of initial gas stripped, as a comparison
to the results presented in Figure \ref{fig5} and Figure \ref{fig6} in \S\ref{results_cumulative}.
At the $25\%$ stripping timescale
for the neutral gas, we find that the median estimated contribution
from tides is $0.35\%$ of the gas mass stripped by that time
(i.e. $0.09\%$ of the total neutral gas mass). Six
of the 69 galaxies ($9\%$) that reach $25\%$ neutral gas stripping
have tidal contributions of over $10\%$ to the mass stripped at that time.
At the $75\%$ stripping timescale of the neutral gas, the median contribution from
tides is $1.4\%$ and 3 of 48 galaxies ($6\%$) have tidal contributions of over $10\%$.
As expected, the ionized gas is somewhat more influenced by tides: at the $25\%$ stripping timescale, the
median tidal contribution is $1.2\%$ and 7 of 79 galaxies ($9\%$) have contributions over $10\%$.
Finally, for the $90\%$ stripping timescale of the ionized gas, the median tidal
contribution is $2.2\%$ and 9 of 57 galaxies ($16\%$) have contributions over $10\%$.

Given the above estimates of the effect of tides on our sample, and
especially considering that they are likely to be overestimates,
we conclude that the results presented in this paper are not significantly
affected by the presence of tides.

\end{document}